\documentclass[aps,prd,superscriptaddress,twocolumn,10pt]{revtex4-1}

\usepackage{graphicx,amsfonts,amsmath,amssymb,amstext}
\usepackage{float,wrapfig}
\usepackage{subfigure, psfrag}
\usepackage{dsfont}
\usepackage{color}
\usepackage{bm}
\usepackage{hyperref}
\hypersetup{
	colorlinks = true,
	linkcolor = blue,
	citecolor = blue
}

\newcommand{\sign}[1]{\,\mbox{sgn}\left({#1}\right)}

\definecolor{purple}{rgb}{0.8,0,0.6}
\definecolor{darkgreen}{rgb}{0.00,0.6,0.00}

\def\mytitle{Non-Hermitian impurities in Dirac systems}

\begin{document}

\title{\mytitle}
\date{December 31, 2019}

\author{P.~O.~Sukhachov}
\email{pavlo.sukhachov@su.se}
\affiliation{Nordita, KTH Royal Institute of Technology and Stockholm University, Roslagstullsbacken 23, SE-106 91 Stockholm, Sweden}

\author{A.~V.~Balatsky}
\email{avb@nordita.org}
\affiliation{Nordita, KTH Royal Institute of Technology and Stockholm University, Roslagstullsbacken 23, SE-106 91 Stockholm, Sweden}
\affiliation{Department of Physics, University of Connecticut, Storrs, CT 06269, USA}

\begin{abstract}
Quasiparticle states in Dirac systems with complex impurity potentials are investigated. It is shown that an impurity site with loss leads to a nontrivial distribution of the local density of states (LDOS). 
While the real part of defect potential induces a well-pronounced peak in the density of states (DOS), the DOS is either weakly enhanced at small frequencies or even forms a peak at the zero frequency for a lattice in the case of non-Hermitian impurity.
As for the spatial distribution of the LDOS, it is enhanced in the vicinity of impurity but shows a dip at a defect itself when the potential is sufficiently strong. The results for a two-dimensional hexagonal lattice demonstrate the characteristic trigonal-shaped profile for the LDOS. The latter acquires a double-trigonal pattern in the case of two defects placed at neighboring sites. The effects of non-Hermitian impurities could be tested both in photonic lattices and certain condensed matter setups.
\end{abstract}

\maketitle

\section{Introduction}
\label{sec:introduction}

Non-Hermitian systems that are governed by non-Hermitian operators~\cite{Moiseyev:book,Bender:book} are the subject of a significant interest.
One of the most studied cases are the $\mathcal{PT}$-symmetric Hamiltonians (see  Refs.~\cite{Bender:2005-rev,Bender:2007-rev,Mostafazadeh:2010-rev,Makris-Musslimani:2011-rev,Konotop-Zezyulin:2016-rev} for reviews). As was first suggested in Refs.~\cite{Bender-Boettcher:1998,Bender-Meisinger:1999}, the combined action of the parity-inversion $\mathcal{P}$ and time-reversal $\mathcal{T}$ symmetries allows for purely real spectra of non-Hermitian operators. Gain and loss for such $\mathcal{PT}$-symmetric systems are compensated.

Recently, the concept of non-Hermitian and, in particular, $\mathcal{PT}$-symmetric systems, spread beyond quantum mechanics to metamaterials and artificial structures. Such a surge of interest is explained by experimental realizations of various non-Hermitian phases in optical waveguides and lattices (for a recent brief comment on the prospects of the non-Hermitian optics, see Ref.~\cite{El-Ganainy-Ozdemir:2019}).
Non-Hermitian Hamiltonians are able to capture certain aspects of open systems with gain and loss. Moreover, the concept of topology is also  applied to the non-Hermitian domain~\cite{Kawabata-Ueda:2019,Kawabata-Sato:2019,Zhou-Lee:2019,Ota-Iwamoto:2019-rev}. For example, non-Hermitian boundary modes and the bulk-boundary correspondence~\cite{Yao-Wang:2018a,Yao-Wang:2018b,Martinez-Torres:2018,Xiong:2018,Lee-Thomale:2018,Borgnia-Slager:2019,Zirnstein-Rosenow:2019} are actively investigated.

Photonic lattices proved to be a particularly fertile ground to investigate topologically nontrivial phases in non-Hermitian systems ~\cite{Rechtsman-Szameit:2013,Hafezi-Taylor:2013,Maczewsky-Szameit:2017,Hassan-Bourennane:2019,Bergholtz-Kunst:rev-2019}.
These lattices can be effectively used also to probe various disorder and transport effects. For example, the Anderson localization was experimentally observed in two-dimensional (2D) photonic lattice in Ref.~\cite{Schwartz-Segev:2007}.
Furthermore, optical waveguides and photonic lattices allow for the realization of the so-called photonic graphene, which is discussed, for instance, in Refs.~\cite{Peleg-Christodoulides:2007,Szameit-Segev:2007,Bahat-Treidel-Segev:2008,Liang-Dong:2011,Kartashov-Torner:2013,Plotnik-Segev:2014,Wu-Hou:2019}. These examples point to the growing focus in artificial topological and Dirac materials.

The notion of Dirac materials became widely spread in modern condensed matter. These materials represent a special class of two- and three-dimensional (3D) systems whose quasiparticles demonstrate relativistic-like behavior. Among them are graphene~\cite{Katsnelson:rev-2007,Geim-Novoselov:rev-2007,CastroNeto-Geim:rev-2009,Geim:rev-2009,Katsnelson:book-2012,Wehling-Balatsky:rev-2014}, topological insulators~\cite{Hasan-Kane:rev-2010,Qi-Zhang:rev-2010,Hasan-Moore:rev-2011,Bernevig:book-2013,Wehling-Balatsky:rev-2014}, 3D Dirac and Weyl semimetals~\cite{Wehling-Balatsky:rev-2014,Yan-Felser:2017-Rev,Hasan-Huang:rev-2017,Armitage-Vishwanath:2017-Rev}, etc. The valence and conduction bands of Dirac materials touch in isolated Dirac points of the Brillouin zone allowing one to apply the Dirac equation for the description of quasiparticle properties. This leads for many interesting effects including, in particular, nontrivial manifestations of disorder.

It is well known that impurities and defects significantly affect the properties of various materials. In particular, they are crucial for most of transport properties and play a key role in semiconductor physics. The manifestations of disorder are not limited to hindering quasiparticle propagation and broadening of spectral lines only. Defects enable new phenomena and effects as well.
The latter include the appearance of additional electronic states known as impurity resonances (for a review, see Refs.~\cite{Balatsky-Zhu:2006,Wehling-Lichtenstein:2009,Skrypnyk-Loktev:rev-2018}). As was noted by A.~M.~Lifshitz~\cite{Lifshitz:1964a,Lifshitz:1964b}, the presence of even a small number of defects can lead to a noticeable spectrum rearrangement near van Hove singularities. In the case of relativistic-like spectrum, the latter correspond to Dirac points. In metals due to a large density of states (DOS) in conventional metals, the modification due to impurity resonances is usually negligible. On the other hand, the DOS in Dirac systems has the characteristic linear (2D) or quadratic (3D) dependence on energy and vanishes in a Dirac point. This results in the creation of the resonance peak near a minimum in the DOS~\cite{Pereira-CastroNeto:2006,Peres-CastroNeto:2006,Skrypnyk-Loktev:2006,Wehling-Wiesendanger:2007,Wehling-Katsnelson:2010,Chen-Haas:2011}.
The other hallmark feature of impurity resonances is the specific pattern of the local DOS (LDOS), whose trigonal form was observed via the scanning tunneling microscopy (STM) in graphene~\cite{Ugeda-Gomez-Rodriguez:2010,Gomes-Manoharan:2012,Jiang-Andrei:2018,Wen-Yan-Shi-Xuan:2019}.

An impurity potential in condensed matter systems is usually taken to be real. It can be either attractive or repulsive.
For example, vacancies and ad-atoms like hydrogen are typical impurities in graphene~\cite{Pereira-CastroNeto:2006,Peres-CastroNeto:2006,Skrypnyk-Loktev:2006,Wehling-Wiesendanger:2007,Yazyev-Helm:2007,Wehling-Katsnelson:2010,Yazyev:2010,Chen-Haas:2011}. Recent interest in non-Hermitian models opens up a new set of questions on the role of non-Hermitian defects in condensed matter setting including photonic and cold atoms platforms. 

The focus of this study is the non-Hermitian effects in Dirac matter, e.g., in photonic graphene. 
Motivated by experimental~\cite{Bartal-Segev:2005,Schwartz-Segev:2007} and recent theoretical~\cite{Wu-Hou:2019,Zhang-Song:2019,Xue-Chong:2019} studies in optical non-Hermitian systems, we investigate the effects of \emph{non-Hermitian defects} in 2D and 3D Dirac systems. In particular, we show how the presence of the imaginary part of defect potential affects the spatial distribution of the LDOS and changes the frequency dependence of the DOS. Our analytical results are supported by tight-binding calculations for a 2D hexagonal lattice. Among the key findings is the redistribution of the DOS for a dissipative potential. Like in the case of usual impurities, the LDOS demonstrates a peak near the defect (continuum model) or trigonal profile (lattice). This is particularly interesting since no well-pronounced peaks appear in the frequency dependence of the DOS in this case. 
In that regard we point the related but different proposal in Ref.~\cite{Zhang-Song:2019}, where the formation of bound states inside the band gap due to a $\mathcal{PT}$-symmetric imaginary potential in strongly coupled bilayer lattices was considered. We also discuss experimental feasibility of non-Hermitian impurity in electronic and photonic systems. 

The paper is organized as follows. In Sec.~\ref{sec:model-general}, we introduce a continuum model and the key details of formalism. The results for the frequency and spatial dependence of the DOS in this model are presented in Sec.~\ref{sec:model-continuum}. Section~\ref{sec:lattice} is devoted to the 2D hexagonal lattice model. Finally, the conclusions are presented in Sec.~\ref{sec:conclusions}. The results for a 3D continuum Dirac Hamiltonian are presented in Appendix~\ref{sec:continuum-3D}. The spatial dependence of the LDOS in the lattice model for an imaginary potential is analyzed in Appendix~\ref{sec:lattice-Im}. Through this study, we set $\hbar=k_{B}=1$.

\section{Continuum Dirac model}
\label{sec:model}

\subsection{General definitions}
\label{sec:model-general}

In this section, we formulate the continuum model of gapless Dirac systems and discuss the effects of a single impurity. The results for a simple tight-binding Hamiltonian of a 2D hexagonal lattice are presented in Sec.~\ref{sec:lattice}.
In the absence of mass terms, the low-energy Dirac Hamiltonian can be rewritten as a direct sum of two Weyl Hamiltonians $H_{4\times4}=H_{2\times2}^{(+)}\oplus H_{2\times2}^{(-)}$. Here
\begin{equation}
\label{Model-H-1}
H_{2\times2}^{(\lambda)} =H_0= \int d^n r\, \psi_{\lambda}^{\dag}(\mathbf{r}) \left[ -i\lambda \hbar v_F \left(\bm{\sigma}\cdot \bm{\nabla}\right) \right]\psi_{\lambda}(\mathbf{r}),
\end{equation}
where $\lambda=\pm$ describes $K$ and $K^{\prime}$ valley in graphene or chirality of Weyl fermions and $n=2$ ($n=3$) in 2D (3D) case. Further, $\bm{\sigma}$ is the vector of the Pauli matrices acting in the pseudospin space and $v_F$ is the Fermi velocity (or its analog in optical lattice setups).

The impurity is described via the following interaction term:
\begin{equation}
\label{Model-H-int}
H_{\rm int} = \int d^n r\, \psi_{\lambda}^{\dag}(\mathbf{r}) U(\mathbf{r}) \psi_{\lambda}(\mathbf{r}).
\end{equation}
We assume the simplest model of spinless, short-range impurities that do not intermix chiralities, i.e., $U(\mathbf{r})= U \delta(\mathbf{r})\mathds{1}_4$. It is important to emphasize that the impurity potential $U$ in the case under consideration is complex. While $\mbox{Im}\left[U\right]<0$ corresponds to a dissipative or lossy defect, $\mbox{Im}\left[U\right]>0$ allows for an amplification or gain. In what follows, we will primarily focus on the case of lossy defects, which can be routinely realized in optical lattices.

To investigate effects of impurity and the formation of impurity resonances, we employ the standard transfer matrix ($T$-matrix) formalism~\cite{Doniach-Sondheimer:book}. Since the only non-Hermitian part of the Hamiltonian is related to the impurity potential, the eigenvalues of the noninteracting Hamiltonian $H_0$ are real. Therefore, one can define the following bare retarded Green's function in the frequency-momentum space:
\begin{equation}
\label{Model-G-R-0}
G_0^{\rm R}(\omega,\mathbf{k}) = \frac{\omega +\lambda v_F(\bm{\sigma}\cdot\mathbf{k})}{\omega^2-v_F^2k^2 +i0\sign{\omega}}.
\end{equation}
The full Green's function in the $T$-matrix approach is built upon the retarded Green's functions of noninteracting quasiparticles and reads as
\begin{equation}
\label{Model-G-R-full}
G(\omega,\mathbf{k},\mathbf{k}^{\prime}) = \delta_{\mathbf{k},\mathbf{k}^{\prime}} G_0^{\rm R}(\omega,\mathbf{k}) + G_0^{\rm R}(\omega,\mathbf{k}) T_{\mathbf{k},\mathbf{k}^{\prime}} G_0^{\rm R}(\omega,\mathbf{k}^{\prime}).
\end{equation}
The $T$-matrix for a local disorder itself is defined as $T_{\mathbf{k},\mathbf{k}^{\prime}}=\delta_{\mathbf{k},\mathbf{k}^{\prime}} T$, where
\begin{equation}
\label{Model-T-def}
T = U + U\sum_{\mathbf{k}} G_0^{\rm R}(\omega,\mathbf{k}) T.
\end{equation}
From the latter equation, it is straightforward to obtain the following $T$-matrix:
\begin{equation}
\label{Model-T-calc}
T = \left[1- U\sum_{\mathbf{k}} G_0^{\rm R}(\omega,\mathbf{k})\right]^{-1}U,
\end{equation}
where $\sum_{\mathbf{k}} =  V_{\rm BZ}^{-1}\int d^nk/(2\pi)^n$ and $V_{\rm BZ}$ is the volume of the Brillouin zone. In the case of the linearized model at hand, $V_{\rm BZ}=\Lambda^n$, where $\Lambda$ is the momentum cutoff. According to Eq.~(\ref{Model-G-R-full}), the poles of the $T$-matrix are also the poles of the full Green's function $G(\omega,\mathbf{k},\mathbf{k}^{\prime})$, which give bound states and resonances.
As long as we consider lossy impurities, the analytical properties of the $T$-matrix remain unchanged. As to the case of defects with gain, it requires a special treatment and will be addressed elsewhere.

Bound states and resonances correspond to sharp features in the frequency and position dependence of the LDOS~\cite{Balatsky-Zhu:2006,Wehling-Lichtenstein:2009,Skrypnyk-Loktev:rev-2018}. The LDOS is defined as
\begin{eqnarray}
\label{2D-Dirac-LDOS-def}
\nu(\omega,\mathbf{r}) = -\frac{1}{\pi} \mbox{Im} \, \mbox{tr}\left[G^{\rm R}(\omega,\mathbf{r},\mathbf{r})\right],
\end{eqnarray}
where the full Green's function in the frequency-coordinate representation is
\begin{equation}
\label{2D-Dirac-LDOS-GR-def}
G^{\rm R}(\omega,\mathbf{r},\mathbf{r}^{\prime}) = G_0^{\rm R}(\omega,\mathbf{r}-\mathbf{r}^{\prime}) + G_0^{\rm R}(\omega,\mathbf{r}) T G_0^{\rm R}(\omega,-\mathbf{r}^{\prime}).
\end{equation}
It is clear that the translation invariance is lost due to the presence of defect. Without the loss of generality, the position of the latter was set to $\mathbf{0}$. The first term in the above equation corresponds to the background LDOS, which is independent on $\mathbf{r}$ and reads as
\begin{eqnarray}
\label{2D-Dirac-LDOS-GR-2D}
\nu_{0} = \frac{|\omega|}{2\pi v_F^2 \Lambda^2} \theta\left(v_F \Lambda -|\omega|\right),
\end{eqnarray}
for a 2D Dirac Hamiltonian. Here we took into account only the contribution of a single valley. Since the potential is insensitive to the valley index, the contribution of the other valley is the same. 

\subsection{Impurity resonances and density of states}
\label{sec:model-continuum}

Let us apply the formalism described in Sec.~\ref{sec:model-general} and consider both frequency and spatial dependence of the DOS. Further, we demonstrate the effects of impurities with complex potential for a 2D Dirac semimetal phase. The results for a 3D Dirac case are presented in Sec.~\ref{sec:continuum-3D}.

\begin{figure*}[!ht]
\begin{center}
\includegraphics[width=0.32\textwidth]{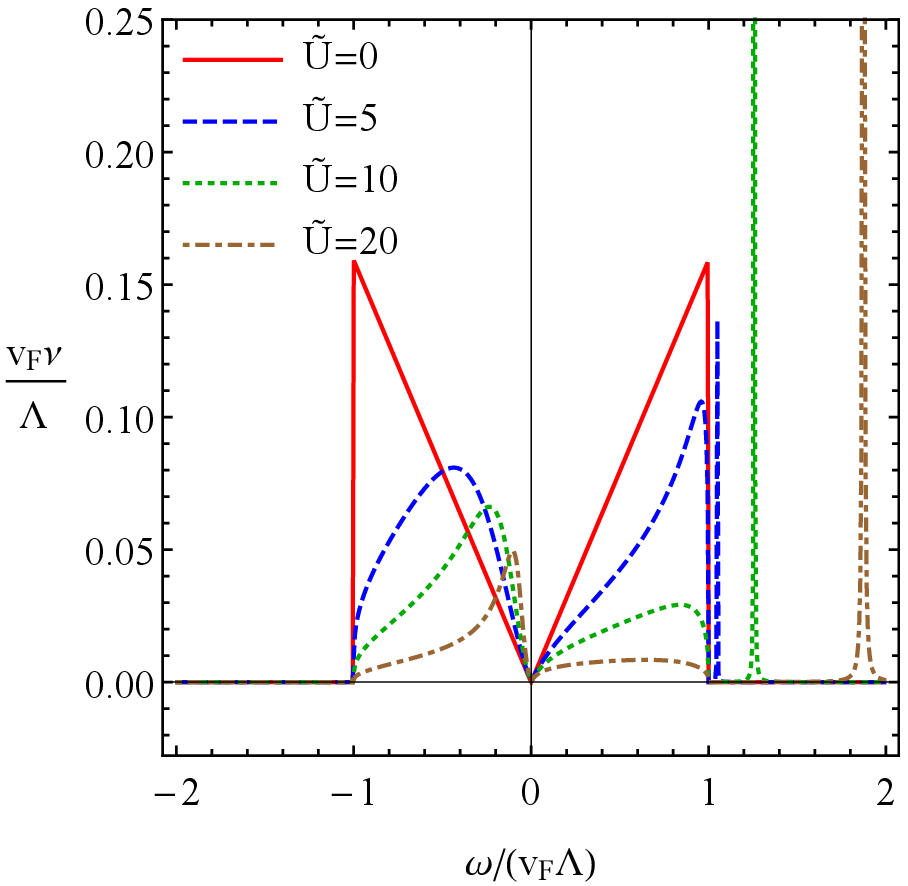}
\hfill
\includegraphics[width=0.32\textwidth]{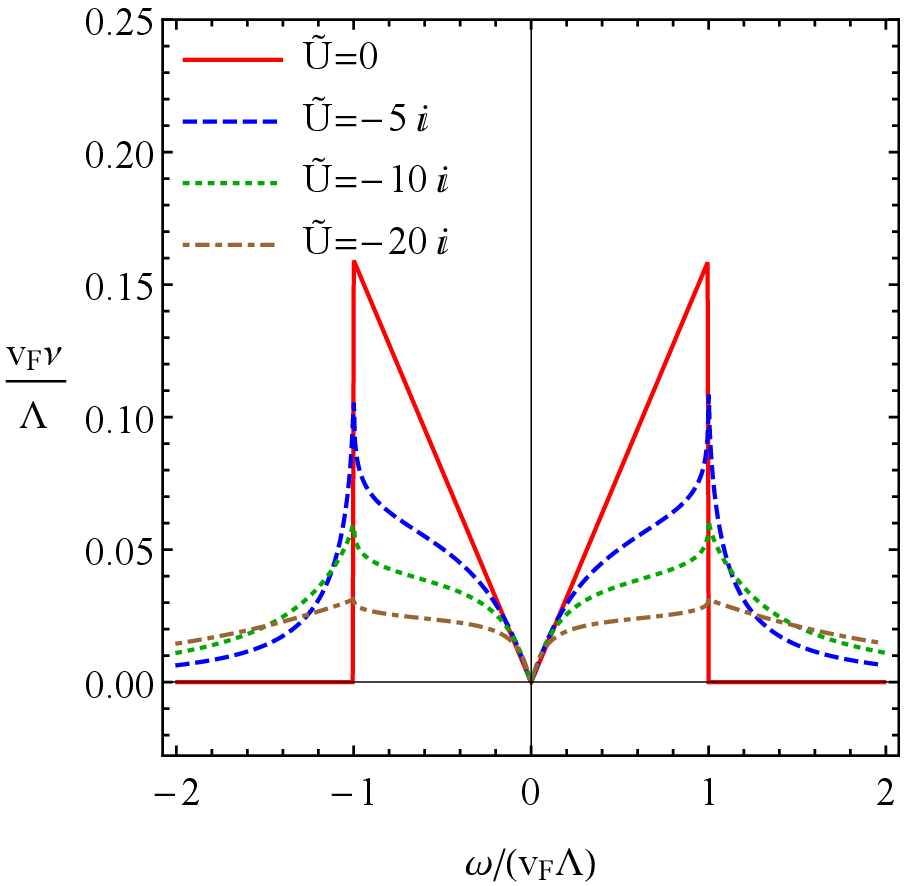}
\hfill
\includegraphics[width=0.32\textwidth]{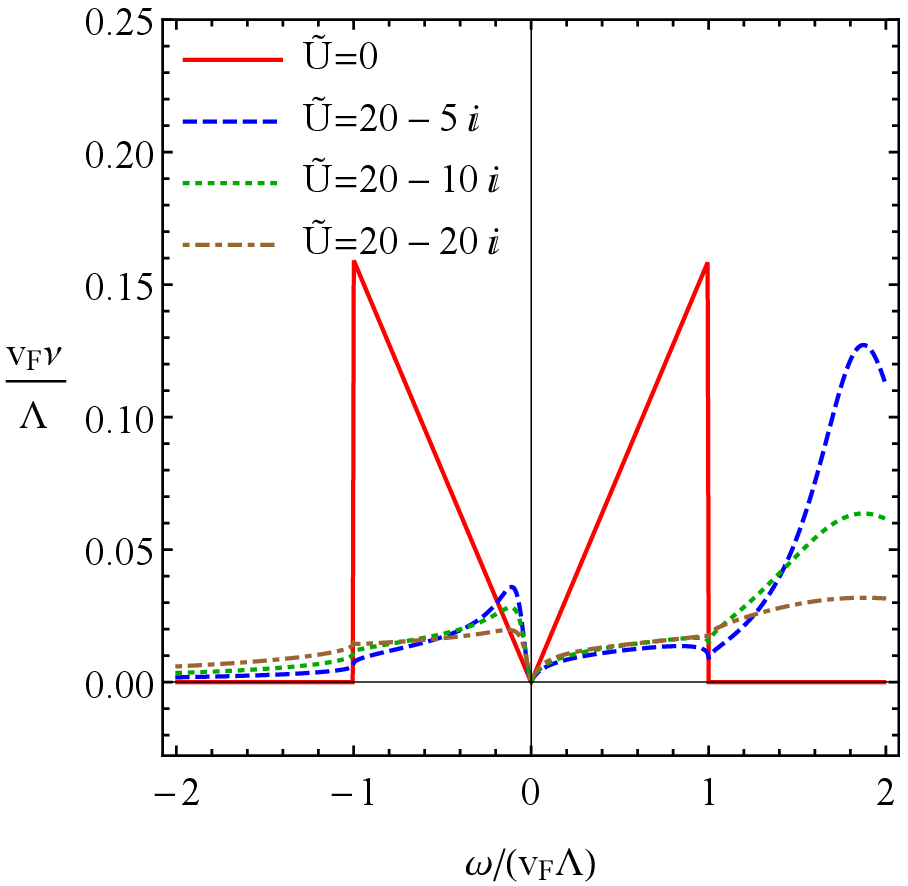}
\end{center}
\hspace{0.01\textwidth}{\small (a)}\hspace{0.32\textwidth}{\small (b)}\hspace{0.32\textwidth}{\small (c)}\\[0pt]
\caption{The LDOS at the impurity site $\nu=\nu(\omega,\mathbf{0})$ for real (panel (a)), imaginary (panel (b)), and complex (panel (c)) impurity potentials as a function of the frequency $\omega$. Red solid, blue dashed, green dotted, and brown dot-dashed lines correspond to different disorder strengths shown in the corresponding legends. In addition, $\tilde{U}=U/(v_F\Lambda)$.
}
\label{fig:2D-Dirac-DOS-omega}
\end{figure*}

The denominator of the $T$-matrix defined in Eq.~(\ref{Model-T-calc}) reads as
\begin{widetext}
\begin{eqnarray}
\label{2D-Dirac-UG-eq}
1-\frac{1}{\Lambda^2} \int \frac{d^2k}{(2\pi)^2}U G_0^{\rm R}(\omega,\mathbf{k}) =1+\frac{U}{\Lambda^2} \frac{1}{4\pi v_F^2} \Bigg[\omega\ln{\left|1-\frac{v_F^2\Lambda^2}{\omega^2}\right|}
&+&i\pi |\omega| \theta\left(v_F\Lambda-|\omega|\right) \Bigg].
\end{eqnarray}
\end{widetext}
Zeroes of this equation determine the resonance and bound states. By using Eqs.~(\ref{2D-Dirac-LDOS-def}) and (\ref{2D-Dirac-LDOS-GR-def}), it is straightforward to determine the LDOS $\nu(\omega,\mathbf{r})$.
We present the latter as a function of the frequency $\omega$ at $\mathbf{r}=\mathbf{0}$ in Fig.~\ref{fig:2D-Dirac-DOS-omega} for a few values of real, imaginary (lossy), and complex potentials. Note that in order to show the presence of the bound states at $|\omega|>\Lambda$, we added a small imaginary part to the denominator of the $T$-matrix, i.e.,  $\theta\left(v_F\Lambda-|\omega|\right)\to \theta\left(v_F\Lambda-|\omega|\right) + 10^{-4} \theta\left(|\omega|- v_F\Lambda\right)$.

Let us first discuss the case of a real potential shown in Fig.~\ref{fig:2D-Dirac-DOS-omega}(a). As it is well known for graphene~\cite{Pereira-CastroNeto:2006,Peres-CastroNeto:2006,Skrypnyk-Loktev:2006,Wehling-Wiesendanger:2007,Wehling-Katsnelson:2010,Chen-Haas:2011}, the presence of impurities with a strong potential reshapes the DOS and allows for the formation of resonance states for $|\omega|<v_F\Lambda$. With the increase of the potential strength, the corresponding peak moves to smaller frequencies and becomes sharper. In addition, there are peaks corresponding to the bound states at large $|\omega|>v_F\Lambda$ that move to larger frequencies with an increase of $U$. (For demonstrative purposes we artificially broadened these peaks).

Further, we turn to one of the key results of this study related to imaginary and complex potentials. Unlike real potential, the lossy one does not lead to the appearance of well-pronounced peaks in the DOS (cf. Figs.~\ref{fig:2D-Dirac-DOS-omega}(a) and \ref{fig:2D-Dirac-DOS-omega}(b)). While, generically, it reduces the DOS at large frequencies, it allows also for a small enhancement at smaller ones. Moreover, the DOS is no longer zero for $|\omega|>v_F \Lambda$. This is explained by the fact that an imaginary potential effectively interchanges real and imaginary parts in the denominator of the $T$-matrix.
In the case of a complex potential, the peak in the DOS is broadened by a nonzero imaginary part of $U$ (see Fig.~\ref{fig:2D-Dirac-DOS-omega}(c)). In addition, due to the redistribution of the DOS caused by $\mbox{Im}\left[U\right]\neq0$, the peak corresponding to the bound states becomes significantly broader.

Note that the case of attractive potential $\mbox{Re}\left[U\right]<0$ is similar, albeit the resonance peaks appear at positive frequencies. In fact, the DOS in this case can be obtained by reflecting $\omega\to-\omega$.

\begin{figure*}[!ht]
\begin{center}
\includegraphics[width=0.32\textwidth]{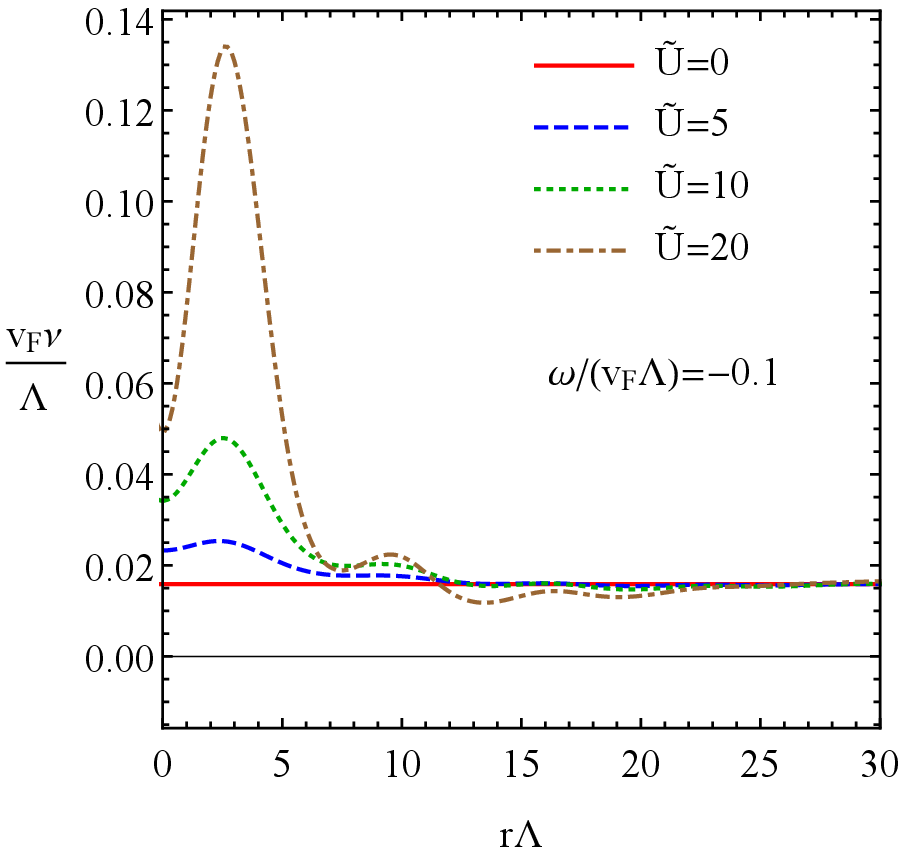}
\hfill
\includegraphics[width=0.32\textwidth]{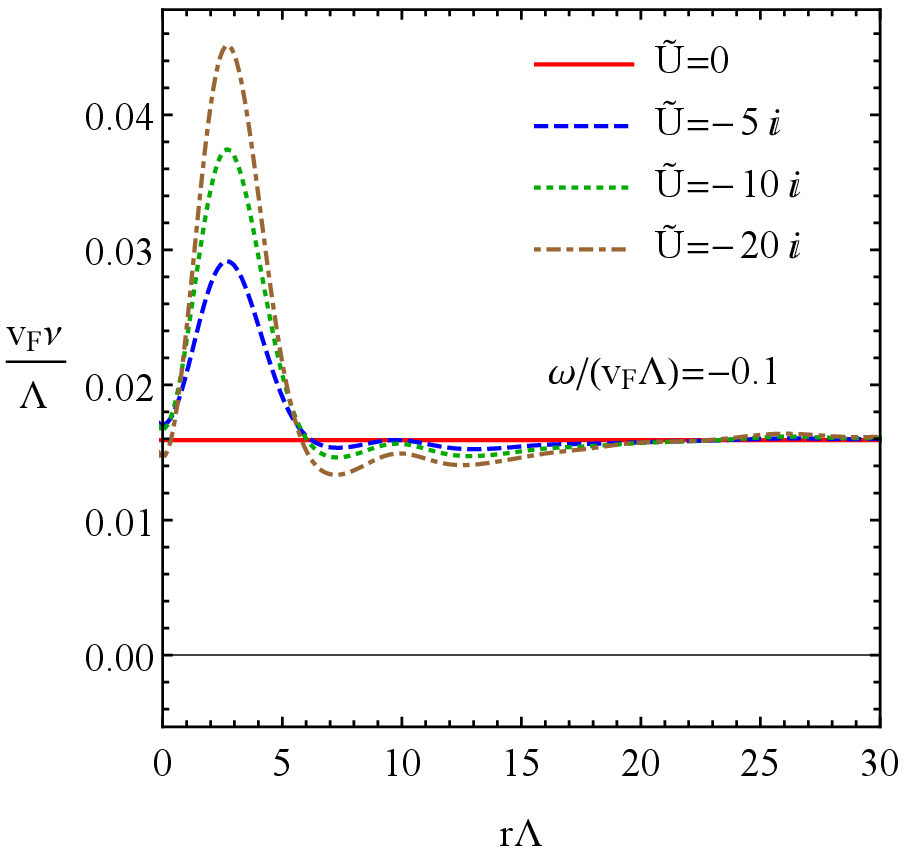}
\hfill
\includegraphics[width=0.32\textwidth]{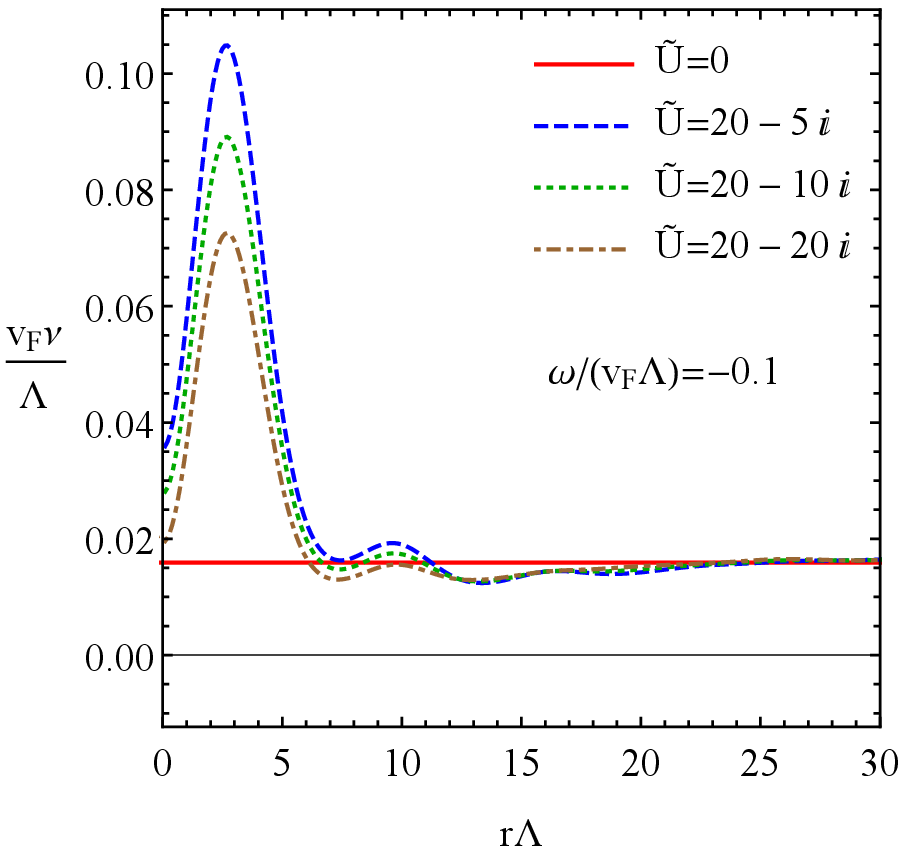}
\end{center}
\hspace{0.01\textwidth}{\small (a)}\hspace{0.32\textwidth}{\small (b)}\hspace{0.32\textwidth}{\small (c)}\\[0pt]
\caption{The LDOS for real (panel (a)), imaginary (panel (b)), and complex (panel (c)) impurity potentials as a function of the radial coordinate $r$. Red solid, blue dashed, green dotted, and brown dot-dashed lines correspond to different disorder strengths shown in the corresponding legends. In addition, we set $\omega=-0.1\,v_F\Lambda$ and $\tilde{U}=U/(v_F\Lambda)$.
}
\label{fig:2D-Dirac-DOS-r}
\end{figure*}

Finally, let us discuss the spatial profiles of the LDOS. In view of the symmetry of the model, we present only the dependence on the radial coordinate $r$. The corresponding results are shown in Figs.~\ref{fig:2D-Dirac-DOS-r}(a), \ref{fig:2D-Dirac-DOS-r}(b), and \ref{fig:2D-Dirac-DOS-r}(c) for a few values of real, imaginary (lossy), and complex potential at a fixed frequency. As one can see, the LDOS distribution has a characteristic profile with a dip at $r=0$ and a well-pronounced peak. Aside from the first peak, there are smaller side peaks, which are related to the Friedel oscillations. It is surprising that the purely imaginary dissipative potential also allows for a similar profile of the LDOS. The deviations of the LDOS due to the impurity are, however, much smaller. The peaks in the LDOS for a complex potential, where, in addition to a real part, there is a negative imaginary one, are suppressed for large $\mbox{Im}\left[U\right]$.

To summarize the key finding in this section, we show that the non-Hermitian dissipative disorder allows for a nontrivial distribution of the DOS. While the well-pronounced resonance peaks in frequency are absent, the spatial dependence of the LDOS is similar to the case of a usual impurity resonance. As is explicitly shown in Appendix~\ref{sec:continuum-3D}, the qualitative results found for the 2D model, remain valid also in 3D.

\section{2D lattice model}
\label{sec:lattice}

\subsection{General definitions}
\label{sec:lattice-model}

To investigate the effects related to the lattice symmetry and finite size as well as to provide a support for the results obtained in the continuum model, let us consider a tight-binding model for a 2D hexagonal lattice. Such a lattice qualitatively describes both real and photonic version of graphene. Hexagonal lattice is defined by three vectors:
\begin{equation}
\label{tb-delta-def}
\bm{\delta}_1 = \left(0, a \right), \,
\bm{\delta}_2 = \left(\frac{\sqrt{3}a}{2},-\frac{a}{2}\right), \,
\bm{\delta}_3 = \left(-\frac{\sqrt{3}a}{2},-\frac{a}{2}\right),
\end{equation}
where $a$ is the lattice constant. The corresponding tight-binding Hamiltonian reads as
\begin{eqnarray}
\label{tb-H-def-ab}
H &=& -t\sum_{\mathbf{n}} \sum_{i=1}^{3} \left[\hat{a}_{\mathbf{n}}^{\dag} \hat{b}_{\mathbf{n} +\bm{\delta}_i} +h.c.\right] + U_A \sum_{\mathbf{l}}\hat{a}_{\mathbf{l}}^{\dag} \hat{a}_{\mathbf{l}} \nonumber\\ 
&+&U_B\sum_{\mathbf{l}^{\prime}}\hat{b}_{\mathbf{l}^{\prime}}^{\dag} \hat{b}_{\mathbf{l}^{\prime}},
\end{eqnarray}
where $t$ is the hopping parameter, the creation (annihilation) operators $\hat{a}_{\mathbf{n}}^{\dag}$ ($\hat{a}_{\mathbf{n}}$) and $\hat{b}_{\mathbf{n}}^{\dag}$ ($\hat{b}_{\mathbf{n}}$) act on the sublattices A and B, respectively, as well as $U_A$ and $U_B$ are the on-site impurity (defect) potentials. Further, the first sum runs over all lattice sites as well as the second and third sums accounts for  defects on the A and B sublattices, respectively. We assume that there are $N$ lattice sites. As for the defects, the cases with one and two impurities will be discussed.

\begin{figure}[!ht]
\begin{center}
\includegraphics[width=0.45\textwidth]{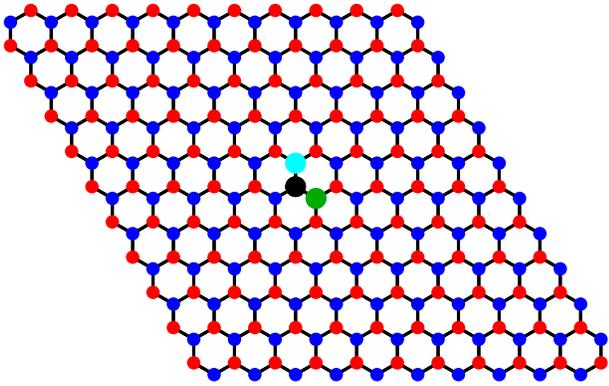}
\end{center}
\caption{Hexagonal lattice with $N=240$ sites ($10\times10$ hexagons) with zigzag boundary conditions. Red and blue dots corresponds to the A and B sublattices, respectively. The black and green dots mark impurity (defect) positions. The cyan dot is the neighbouring site, which will be used to present the frequency dependence of the LDOS.
}
\label{fig:tb-lattice}
\end{figure}

In what follows, we consider the case of a lattice with $N=240$ sites formed by $10\times10$ hexagons. We checked that the key results remain qualitatively the same for a small lattice with $N=48$ sites ($4\times4$ hexagons). In the latter case, the distortions due to boundary effects are more noticeable, however. The lattice together with the positions of the impurities is shown in Fig.~\ref{fig:tb-lattice}. Note that we use the zigzag boundary conditions.
To avoid the effects of boundaries as much as possible, the impurities are placed close to the center of the lattice.
The positions of impurities are marked by the black and green dots in Fig.~\ref{fig:tb-lattice}.

The LDOS for a lattice model is defined as
\begin{eqnarray}
\label{tb-DOS-def}
&&\nu(\omega, \mathbf{r}_i) = \sum_{j} \psi_{j}^{\dag}(\mathbf{r}_i)\psi_j(\mathbf{r}_i)  \delta_{\tilde{\Gamma}}\left[\omega - \mbox{Re}(\epsilon_j)\right] \nonumber\\
&&= \sum_{j} \frac{\psi_{j}^{\dag}(\mathbf{r}_i)\psi_j(\mathbf{r}_i)}{\pi} \frac{\Gamma_0-\mbox{Im}(\epsilon_j)}{\left[\omega - \mbox{Re}(\epsilon_j)\right]^2 +\left[\Gamma_0-\mbox{Im}(\epsilon_j)\right]^2}. \nonumber\\
\end{eqnarray}
Here $\mathbf{r}_i$ denotes the position of the site where the LDOS is calculated and $\sum_{j}$ runs over all eigenvalues $\epsilon_j$. Further, $\psi_{j}^{\dag}(\mathbf{r}_i)$ is the normalized eigenvector corresponding to $\epsilon_j$ and $\tilde{\Gamma} = \Gamma_0-\mbox{Im}(\epsilon_j)$.
Note that we introduced a Lorentzian broadening of the $\delta$-function by $\Gamma_0$. This background broadening corresponds, e.g., to temperature in condensed matter setups or overall loss for a photonic lattice. The latter could be also modeled by adding on-site dissipative potential for all lattice sites.
In our numerical simulations, it is assumed that $\Gamma_0=0.05\,|t|$. As we will demonstrate below, the presence of lossy impurities allows for $\mbox{Im}\left[\epsilon_j\right]<0$, which will also contribute to the broadening.

As in the case of the continuum model, we consider the spatial distribution of the LDOS. In the model at hand, the LDOS is discrete and localized at lattice sites. Therefore, for presentation purposes, we introduce the spatial broadening by the normal distribution and define the following continuous LDOS:
\begin{eqnarray}
\label{tb-DOS-cont}
\nu(\omega, \mathbf{r})= \sum_{i}\frac{1}{\sqrt{2\pi} r_{\rm SD}} e^{-\left(\mathbf{r}-\mathbf{r}_i\right)^2/(2a^2 r_{\rm SD}^2)} \nu(\omega, \mathbf{r}_i),
\end{eqnarray}
where $r_{\rm SD}$ is the standard deviation. We use $r_{\rm SD}=0.3$ in our numerical calculations.

\subsection{Hexagonal lattice with a single impurity}
\label{sec:lattice-DOS}

In this subsection, we present the LDOS as a function of $\omega$ and coordinate for the lattice with a single impurity $U_A=U$ and $U_B=0$. The position of the latter is marked by the black dot in Fig.~\ref{fig:tb-lattice}.

\begin{figure*}[!ht]
\begin{center}
\includegraphics[width=0.475\textwidth]{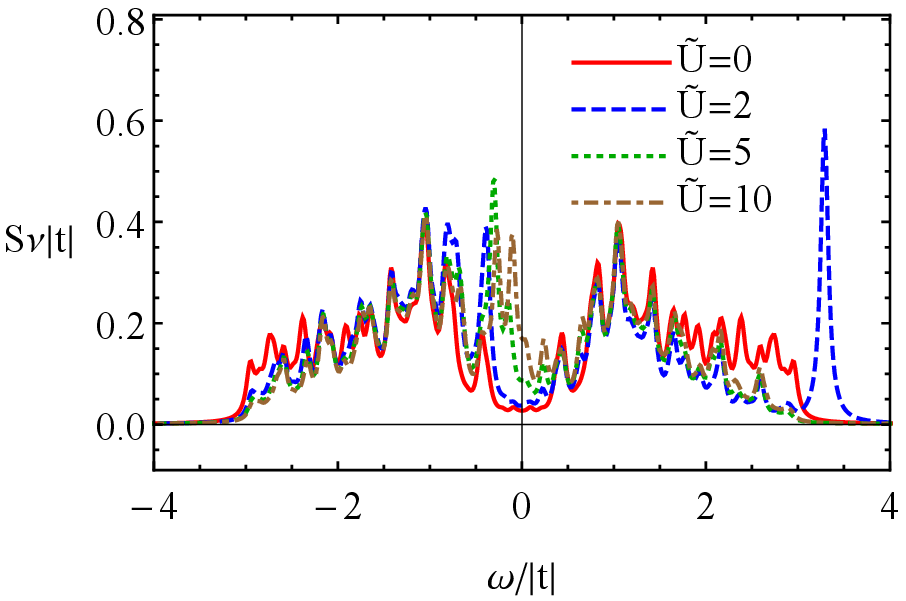}
\hfill
\includegraphics[width=0.475\textwidth]{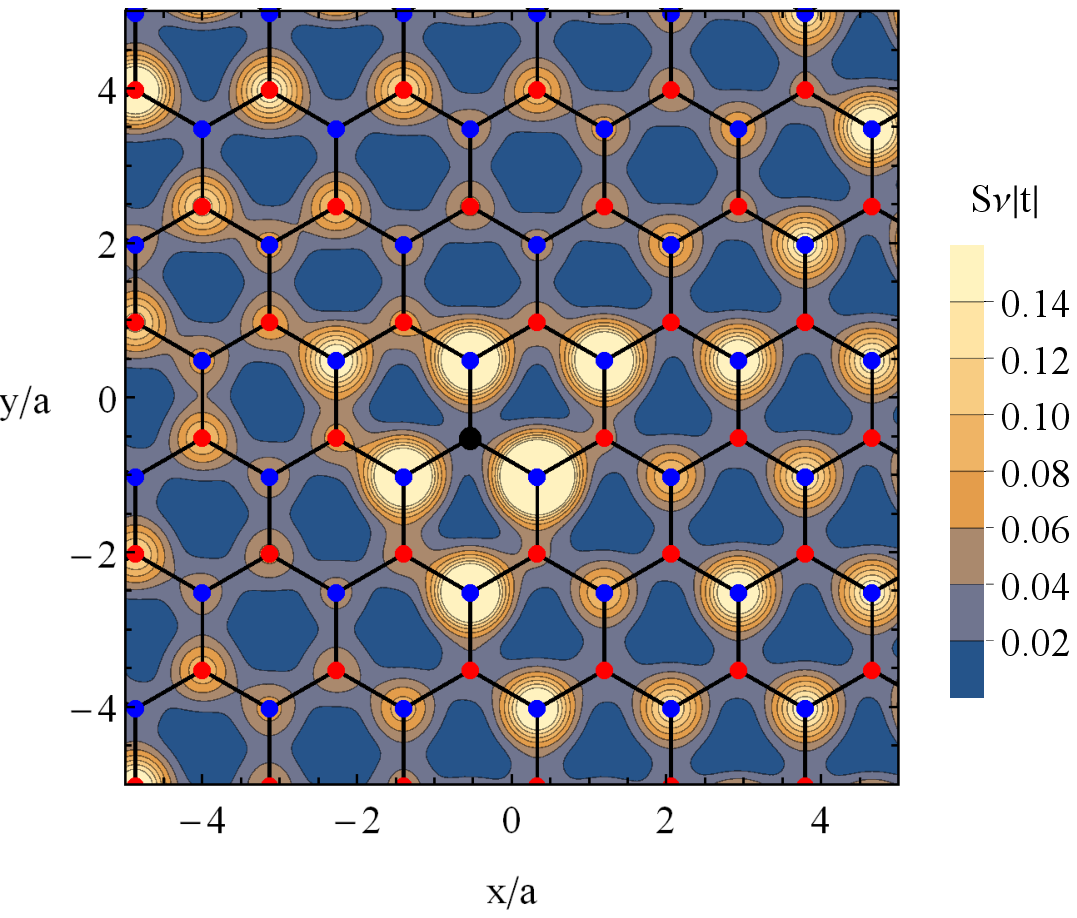}
\end{center}
\caption{Left panel: The LDOS at the neighboring site (see the cyan dot in Fig.~\ref{fig:tb-lattice}) for real impurity potentials as a function of the frequency $\omega$. Red solid, blue dashed, green dotted, and brown dot-dashed lines correspond to different disorder strengths shown in the corresponding legend.
Right panel: The spatial distribution of the LDOS for $\tilde{U}=10$ and $\omega=0$. The position of a single impurity is marked by the black dot.
In both panels, $\tilde{U}=U/|t|$.
}
\label{fig:lattice-DOS-Re}
\end{figure*}

\begin{figure*}[!ht]
\begin{center}
\includegraphics[width=0.475\textwidth]{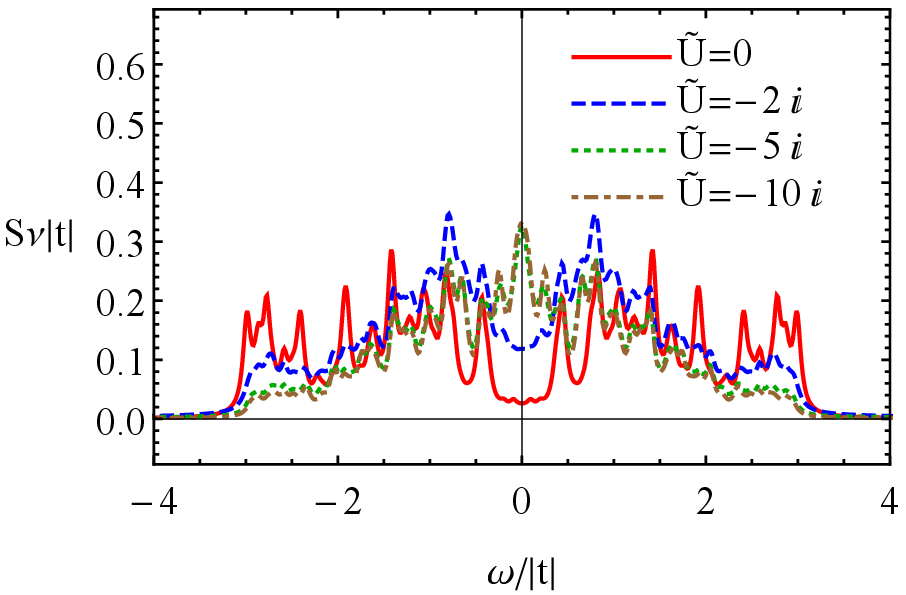}
\hfill
\includegraphics[width=0.475\textwidth]{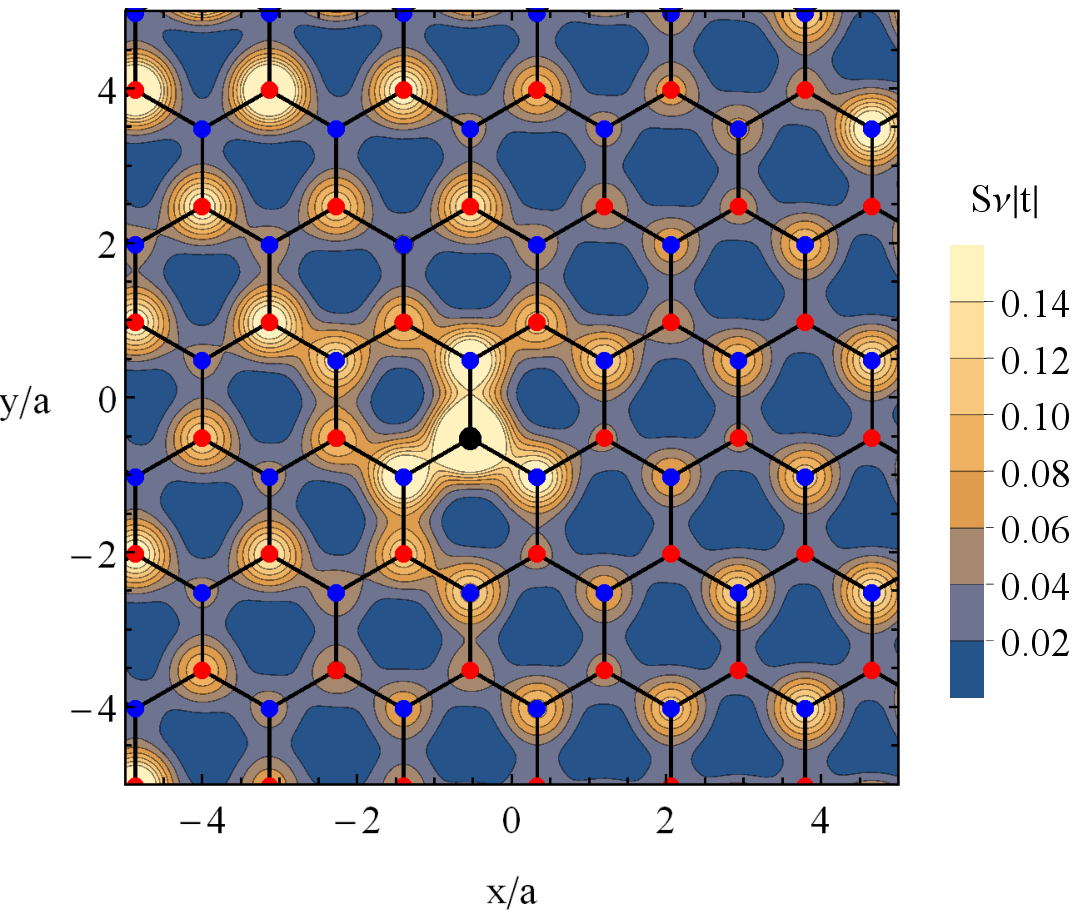}
\end{center}
\caption{Left panel: The LDOS at the neighboring site (see the cyan dot in Fig.~\ref{fig:tb-lattice}) for imaginary impurity potentials as a function of the frequency $\omega$. Red solid, blue dashed, green dotted, and brown dot-dashed lines correspond to different disorder strengths shown in the corresponding legend.
Right panel: The spatial distribution of the LDOS for $\tilde{U}=-2i$ and $\omega=0$. The position of a single impurity is marked by the black dot.
In both panels, $\tilde{U}=U/|t|$.
}
\label{fig:lattice-DOS-Im}
\end{figure*}

\begin{figure*}[!ht]
\begin{center}
\includegraphics[width=0.475\textwidth]{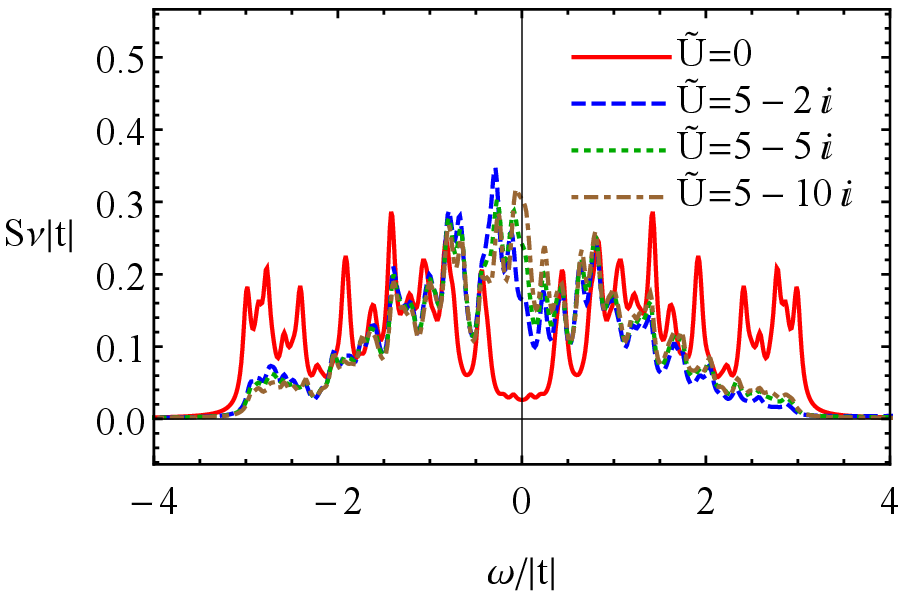}
\hfill
\includegraphics[width=0.475\textwidth]{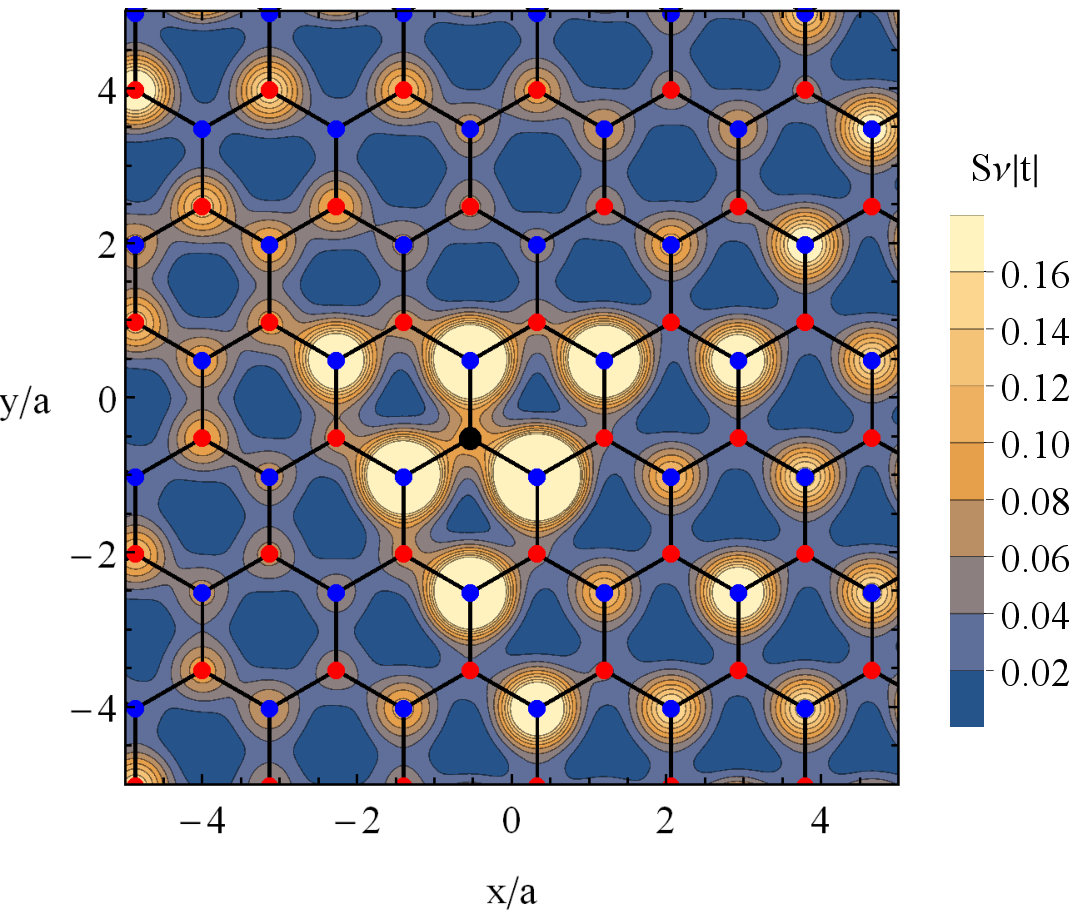}
\end{center}
\caption{Left panel: The LDOS at the neighboring site (see the cyan dot in Fig.~\ref{fig:tb-lattice}) for complex impurity potentials as a function of the frequency $\omega$. Red solid, blue dashed, green dotted, and brown dot-dashed lines correspond to different disorder strengths shown in the corresponding legend.
Right panel: The spatial distribution of the LDOS for $\tilde{U}=5-10i$ and $\omega=0$. The position of a single impurity is marked by the black dot.
In both panels, $\tilde{U}=U/|t|$.
}
\label{fig:lattice-DOS-Re-Im}
\end{figure*}

The dependence of the LDOS on $\omega$ at the neighboring site, which is marked by the cyan dot in Fig.~\ref{fig:tb-lattice}, is shown in the left panels of Figs.~\ref{fig:lattice-DOS-Re}, \ref{fig:lattice-DOS-Im}, and \ref{fig:lattice-DOS-Re-Im} for real, lossy, and complex $U$. Here $S$ is the surface area of the lattice used to normalize the wave functions. Further, we present the spatial distribution of the LDOS at $\omega=0$ at a few values of the impurity potential in the right panels of Figs.~\ref{fig:lattice-DOS-Re}, \ref{fig:lattice-DOS-Im}, and \ref{fig:lattice-DOS-Re-Im}. 

Let us start with the case of real potential. In agreement with the results for the continuum model (see Sec.~\ref{sec:model-continuum}) and previous studies~\cite{Wehling-Wiesendanger:2007,Wehling-Katsnelson:2010,Chen-Haas:2011}, the real repulsive potential leads to the resonance peak in the DOS at small $\omega$ (see the left panel of Fig.~\ref{fig:lattice-DOS-Re}) and allows for the formation of the trigonal-shaped LDOS structure in the vicinity of the impurity presented in the right panel of Fig.~\ref{fig:lattice-DOS-Re}. The peaks in the left panel of Fig.~\ref{fig:lattice-DOS-Re} flatten with increasing the lattice size and performing averaging over the neighboring sites. Note also that the LDOS is accumulated primarily at B-sublattice sites if the defect is on an A-sublattice. In addition, the lattice results suggest that the LDOS at the impurity site itself should be much lower than its neighbourhood.

In agreement with our findings for the continuum model in Sec.~\ref{sec:model-continuum}, the presence of the purely imaginary potential also allows for additional features in the LDOS. In particular, as one can see from the left panel of Fig.~\ref{fig:lattice-DOS-Im}, the DOS at small $\omega$ is enhanced and even form a peak at $\omega=0$ for a strong potential. As for the spatial distribution of the LDOS, it is enhanced at the impurity site at relatively small values of $\mbox{Im}\left[U\right]$, albeit acquires the trigonal shape similar to that in the right panel of Fig.~\ref{fig:lattice-DOS-Re} when $\mbox{Im}\left[U\right]$ becomes large. For a detailed discussion on the corresponding dependence of the LDOS, see Appendix~\ref{sec:lattice-Im}. Finally, the results in Fig.~\ref{fig:lattice-DOS-Re-Im} show that the imaginary part of the impurity potential could enhance the resonance peak for a complex potential but does not modify much the spatial pattern of the LDOS and the position of the peaks.

Thus, in agreement with our finding for the continuum model, the results for the hexagonal lattice also demonstrate the appearance of the distinctive signature in the LDOS for a dissipative potential. Depending on the magnitude of the potential, one can observe an enhancement of the LDOS at the impurity site or conventional trigonal patter of the LDOS. The evolution of LDOS for different values of imaginary potential is also presented in more details in Appendix~\ref{sec:lattice-Im}.

\subsection{Hexagonal lattice with two impurities}
\label{sec:lattice-DOS-2}

\begin{figure*}[!ht]
\begin{center}
\includegraphics[width=0.475\textwidth]{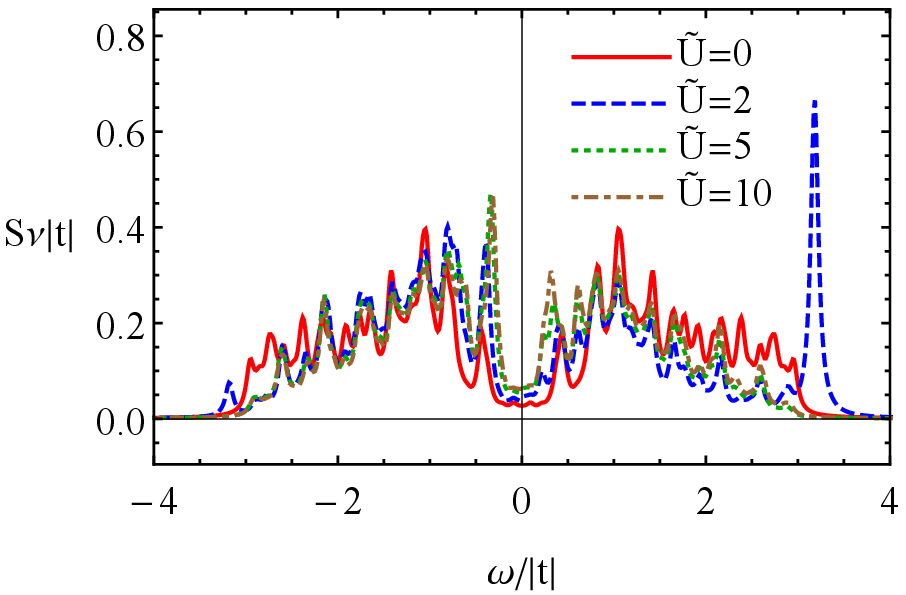}
\hfill
\includegraphics[width=0.475\textwidth]{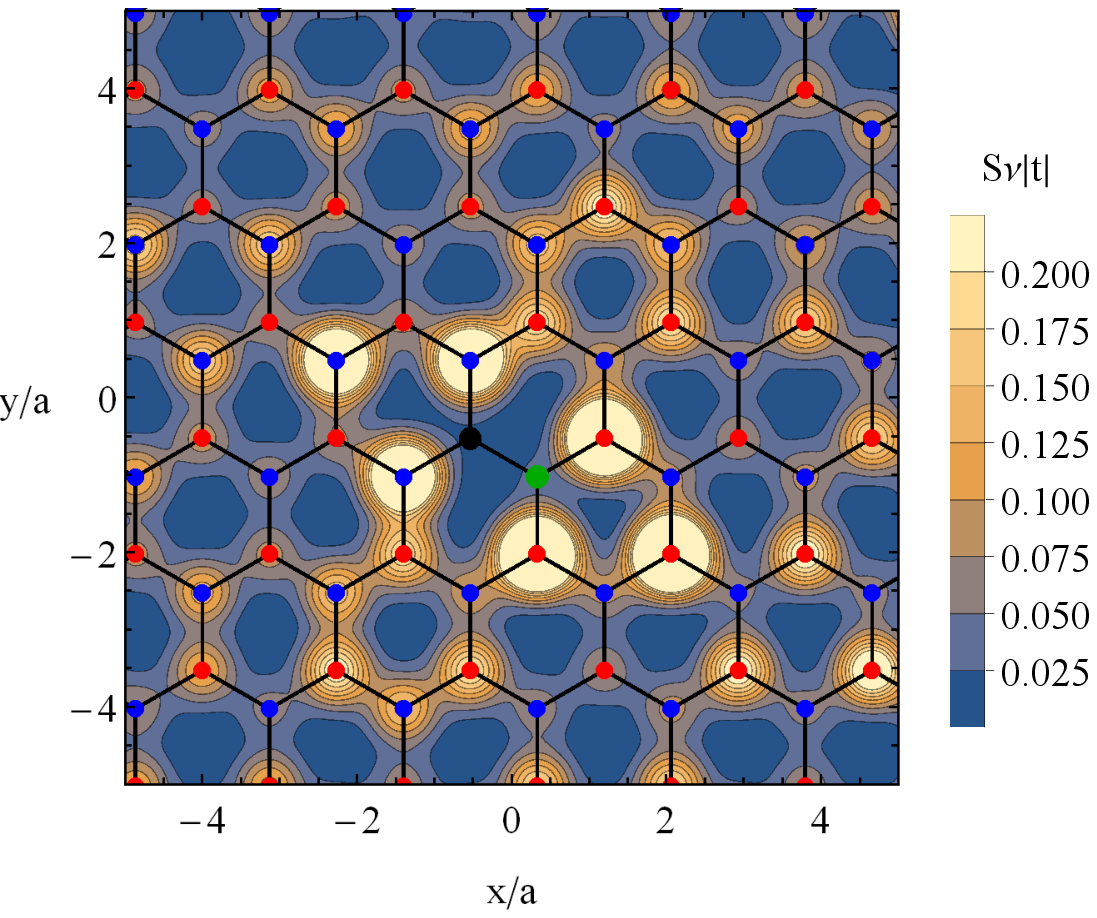}
\end{center}
\caption{Left panel: The LDOS at the neighboring site (see the cyan dot in Fig.~\ref{fig:tb-lattice}) for real impurity potentials as a function of the frequency $\omega$. The values of the potential shown in the legend correspond to the site denoted by a black point.
Right panel: The spatial distribution of the LDOS at $\omega=0$ and $\tilde{U}=10$ for the impurity marked by the black dot.
In both panels, $\tilde{U}=U/|t|$ and it is assumed that the potential at the other defect (green point) is the same in absolute value albeit has an opposite sign of the real part.
}
\label{fig:lattice-DOS-2-Re}
\end{figure*}

\begin{figure*}[!ht]
\begin{center}
\includegraphics[width=0.475\textwidth]{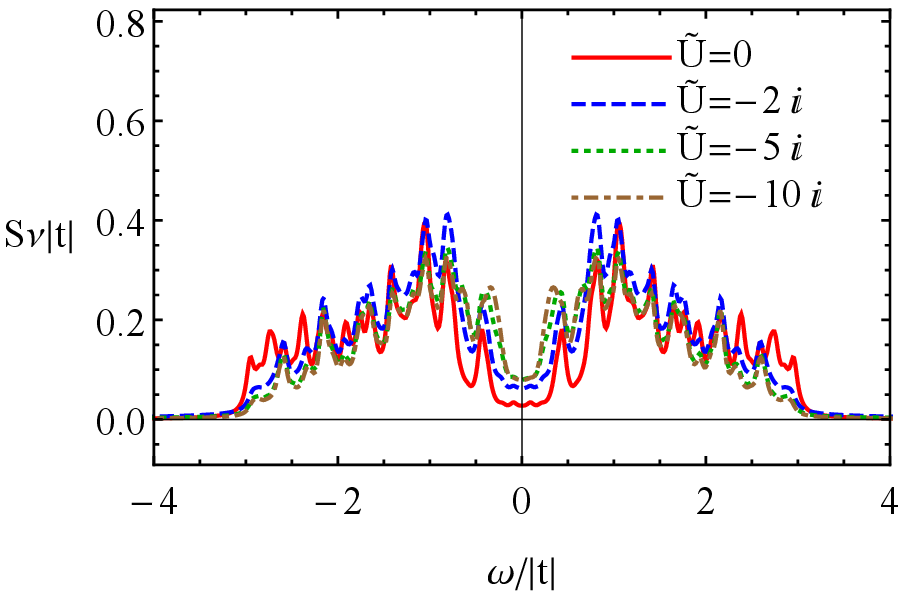}
\hfill
\includegraphics[width=0.475\textwidth]{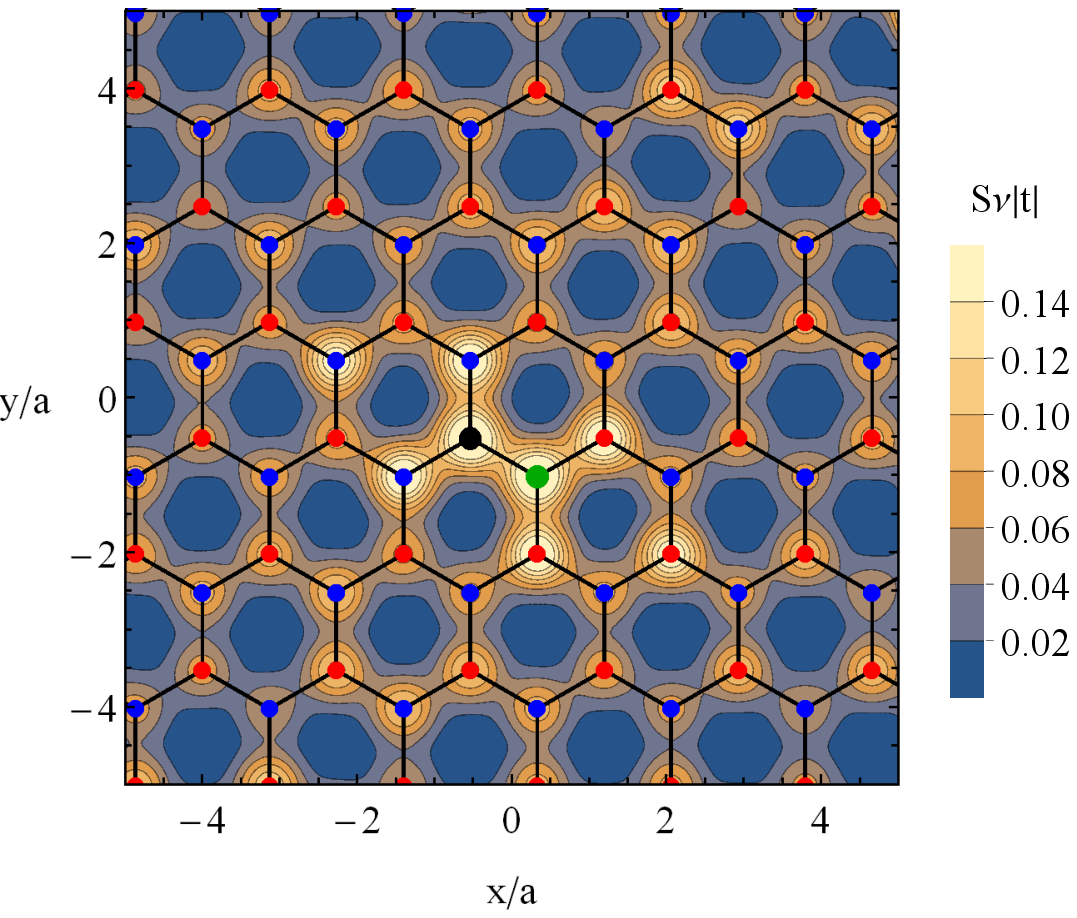}
\end{center}
\caption{Left panel: The DOS at the neighboring site (see the cyan dot in Fig.~\ref{fig:tb-lattice}) for imaginary impurity potentials as a function of the frequency $\omega$. The values of the potential shown in the legend correspond to the site denoted by a black point.
Right panel: The spatial distribution of the LDOS at $\omega=0$ and $\tilde{U}=-2i$ for the impurity marked by the black dot.
In both panels, $\tilde{U}=U/|t|$ and it is assumed that the potential at the other defect (green point) is the same in absolute value albeit has an opposite sign of the real part.
}
\label{fig:lattice-DOS-2-Im}
\end{figure*}

\begin{figure*}[!ht]
\begin{center}
\includegraphics[width=0.475\textwidth]{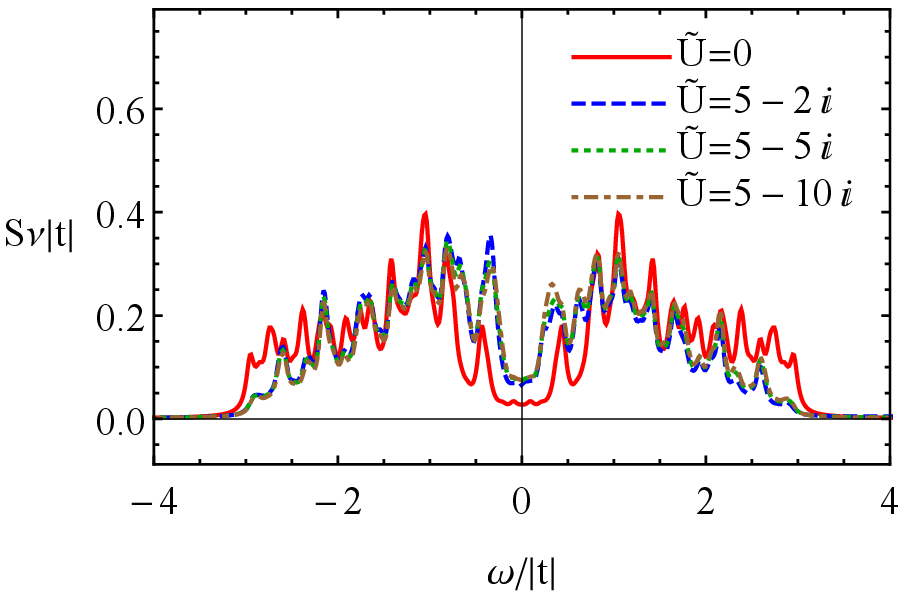}
\hfill
\includegraphics[width=0.475\textwidth]{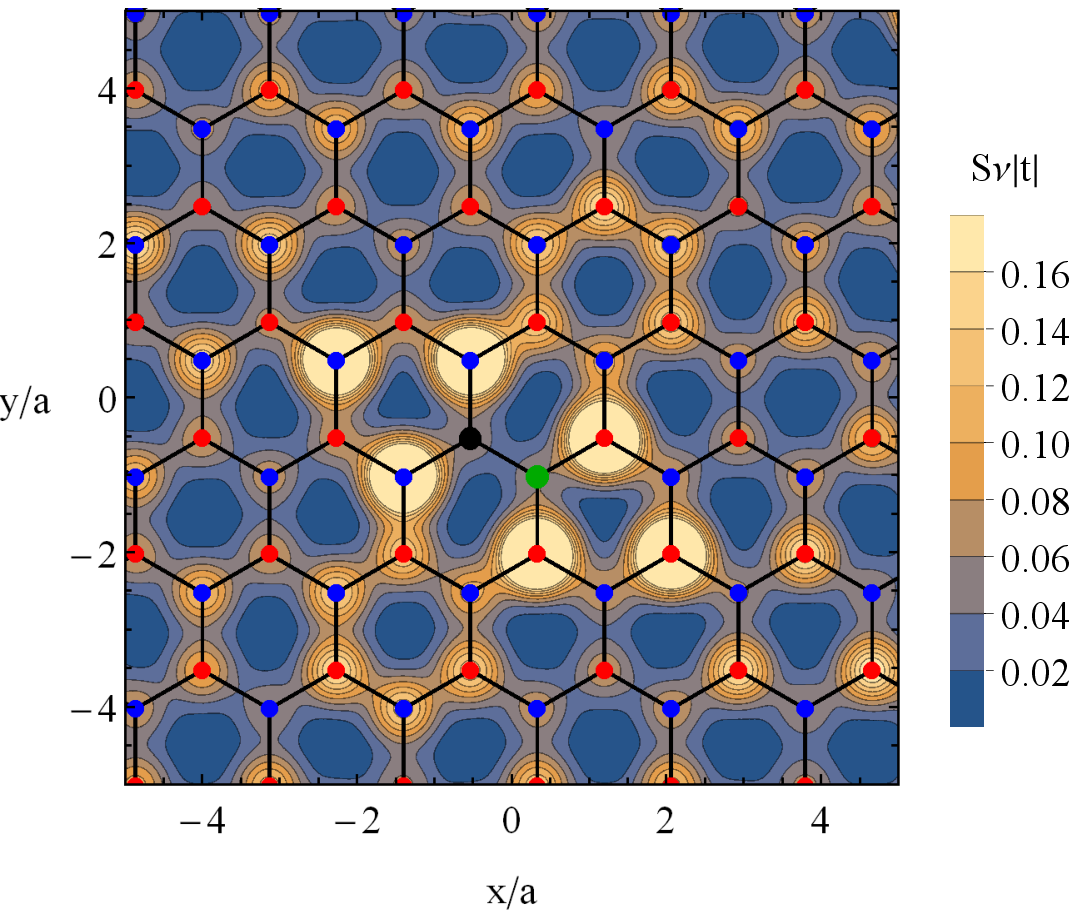}
\end{center}
\caption{Left panel: The DOS at the neighboring site (see the cyan dot in Fig.~\ref{fig:tb-lattice}) for complex impurity potentials as a function of the frequency $\omega$. The values of the potential shown in the legend correspond to the site denoted by a black point.
Right panel: The spatial distribution of the LDOS at $\omega=0$ and $\tilde{U}=5-10i$ for the impurity marked by the black dot.
In both panels, $\tilde{U}=U/|t|$ and it is assumed that the potential at the other defect (green point) is the same in absolute value albeit has an opposite sign of the real part.
}
\label{fig:lattice-DOS-2-Re-Im}
\end{figure*}

In this subsection, we consider the case of two impurities placed on the neighboring sites. Their position is marked by the black and green dots in Fig.~\ref{fig:tb-lattice}. For the sake of brevity, let us investigate only the most interesting case in which the sign of the real part of the impurity potential is opposite for two defects, i.e., $\mbox{Re}\left[U_A\right]=-\mbox{Re}\left[U_B\right]=\mbox{Re}\left[U\right]$ and $\mbox{Im}\left[U_A\right]=\mbox{Im}\left[U_B\right]=\mbox{Im}\left[U\right]$. In particular, this corresponds to a dipole-like impurity where one of the sites is repulsive ($\mbox{Re}\left[U\right]>0$) and the other is attractive ($\mbox{Re}\left[U\right]<0$).
The corresponding results for the LDOS as a function of $\omega$ and $r$ are shown in Figs.~\ref{fig:lattice-DOS-2-Re}, \ref{fig:lattice-DOS-2-Im}, and \ref{fig:lattice-DOS-2-Re-Im} for real, imaginary, and complex potentials.

It is interesting that while the resonance peaks do appear in the LDOS for real impurity potential, the LDOS at $\omega\to0$ is only trivially shifted (see the left panel of Fig.~\ref{fig:lattice-DOS-2-Re}) but becomes asymmetric at $\omega\neq0$. Therefore, in order to probe the nontrivial distribution of the LDOS, a small but nonzero frequency should be used. In particular, we obtained a double-trigonal structure of the LDOS shown in the right panel of Fig.~\ref{fig:lattice-DOS-2-Re} at $\omega=0.3\,|t|$. Note also that the LDOS is accumulated at B-sublattice (A-sublattice) sites near the impurity placed at the A-type (B-type) site.

As in the case of a single impurity, the purely dissipative impurity potentials for two impurities also affects the DOS. As one can see from the left panel in Fig.~\ref{fig:lattice-DOS-2-Im}, the LDOS at $\omega\to0$ is enhanced. As in the case of a single impurity, the shape of the spatial profile of the LDOS depends on the magnitude of the potential. For a relatively small $|U|\sim |t|$, the LDOS in enhanced on the impurity sites. This LDOS quickly acquires double-trigonal shape for large $|U|$, however. (For the details of the LDOS evolution, see Appendix~\ref{sec:lattice-Im}.) The results for the dipole potential with a nonzero dissipative part are shown in Figs.~\ref{fig:lattice-DOS-2-Im} and \ref{fig:lattice-DOS-2-Re-Im}. There is an enhancement of the DOS at small $\omega$ due to the presence of the imaginary part of the potential as well as asymmetry related to $\mbox{Re}\left[U\right]$.

\section{Conclusions}
\label{sec:conclusions}

In this study, we investigated the effects of non-Hermitian defects on the properties of Dirac systems. By using continuum and lattice models, we found that, similarly to conventional impurities described by real potentials, the purely imaginary lossy defects also allow for a nontrivial spatial distribution of the LDOS. The latter is particularly interesting since the corresponding potential does not lead to noticeable peaks in the frequency profile of the DOS.

In agreement with the previous literature, we find the resonance peaks to  occur in the frequency profile of the DOS for a real impurity potential for a  continuum model of 2D and 3D Dirac materials. The position of these peaks is controlled by the potential strength $|U|$. In particular, the peaks move to smaller frequencies with the rise of $|U|$. The impurity resonance is manifested also in the spatial profile of the LDOS, where a noticeable peak occurs in the vicinity of the impurity.

We found that a similar profile of the LDOS appears also for a purely lossy potential. The dependence on the frequency is, however, different. Instead of well-pronounced peaks, the DOS is only slightly enhanced at small frequencies and diminishes for larger ones. Also, the DOS becomes nonzero even for frequencies above the cutoff.
In the case of complex impurity potentials with both real and imaginary parts, the enhancement of the DOS above the cutoff broadens the peaks of bound states at large frequencies, which could make their observation easier. The corresponding signal could be, however, hindered due to possible overlap with other bands.

The numerical results for the 2D hexagonal lattice model with a single impurity support the analytical analysis. In particular, the resonance peaks in the LDOS on a neighboring site appear for a real impurity potential at small frequencies. Moreover, the DOS is enhanced at $\omega=0$ for a purely lossy defect and an impurity with a complex potential. In all these cases, the presence of a defect is manifested in the characteristic enhanced trigonal-shaped LDOS around the impurity when the potential is sufficiently strong. The observation of such a distribution of the LDOS for a lossy site would be a definitive signature of the non-Hermitian defect state.

In the case of a dipole impurity, where the real parts of the defect potentials have opposite signs on neighboring defects, the LDOS as a function of frequency as well as the spatial distribution of the LDOS qualitatively differ from the case of a single defect. In particular, while a resonance peak occurs at small frequencies for a real potential, the DOS at $\omega=0$ is only trivially shifted. The same is also valid for imaginary and complex potential. It is found that the spatial distributions of the LDOS for real, imaginary, and complex impurity potentials have a characteristic double-trigonal form for large $|U|$. In addition, we checked that the qualitative features related to the defect states persist also for smaller lattices, where, however, the effects of boundaries become important.

Let us also briefly discuss the experimental feasibility of our setup. Defects of arbitrary configuration can be straightforwardly realized on-demand in photonic lattices. For example, the impurity potential is directly determined by the size of waveguides. A lossy potential could be achieved by introducing disorder inside the waveguide and/or by not writing it completely (i.e., by leaving regions in the waveguide filled with a background material). Thus, we believe that the model setup of this study can be straightforwardly realized in optical experiments by using a hexagonal photonic lattice. We can speculate that the lossy potential could be also modeled in solids by introducing a sink for electrons (e.g., by employing a point contact). 

\begin{acknowledgments}
We are grateful to M.~Bourennane, W.~Cherifi, and S.~Bandyopadhyay for useful discussions.
This work was supported by the VILLUM FONDEN via the Centre of Excellence for Dirac Materials (Grant No. 11744), the European Research Council under the European Unions Seventh Framework Program Synergy HERO, and the Knut and Alice Wallenberg Foundation KAW 2018.0104.
\end{acknowledgments}

\newpage
\appendix

\begin{widetext}

\section{3D continuum Dirac model}
\label{sec:continuum-3D}

In this section, we consider impurity effects in the three-dimensional (3D) continuum model. The corresponding Hamiltonian is given in Eq.~(\ref{Model-H-1}) in the main text. By using the transfer matrix ($T$-matrix) approach described in Sec.~\ref{sec:model-general}, the resonance and bound states can be identified. In particular, they are given by the poles of the $T$ matrix. In the case of 3D Dirac Hamiltonian, the corresponding characteristic equation reads as
\begin{eqnarray}
\label{3D-Dirac-UG}
&&1-\sum_{\mathbf{k}}U G_0^{\rm R}(\omega,\mathbf{k})= 1-\frac{1}{\Lambda^3}\int \frac{d^3k}{(2\pi)^3}U G_0^{\rm R}(\omega,\mathbf{k})
= 1-\frac{U}{\Lambda^3} \int \frac{d^3k}{(2\pi)^3} \frac{\omega +v_F(\bm{\sigma}\cdot\mathbf{k})}{\omega^2-v_F^2k^2 +i0\sign{\omega}} \nonumber\\
&&=1-\frac{U}{\Lambda^3} \frac{\omega}{2\pi^2} \int_0^{\Lambda} dk\,k^2 \left[\mbox{v.p.} \frac{1}{\omega^2-v_F^2k^2} - i\pi \sign{\omega} \delta\left(\omega^2 -v_F^2k^2\right) \right] \nonumber\\
&&= 1-\frac{U}{\Lambda^3} \frac{\omega}{2\pi^2 v_F^3} \left[-v_F\Lambda \omega + \frac{\omega^2}{2} \ln{\left|\frac{\omega+v_F\Lambda}{\omega-v_F\Lambda}\right|}
-i\frac{\pi}{2} \omega^2 \theta\left(v_F^2\Lambda^2-\omega^2\right) \right].
\end{eqnarray}
Here $U$ is the impurity potential, $G_0^{\rm R}(\omega,\mathbf{k})$ is the bare retarded Green's function given in Eq.~(\ref{Model-G-R-0}), $\Lambda$ is the momentum cutoff, $v_F$ is the Fermi velocity, $\mbox{v.p.}$ stands for the principal value, and $\theta(x)$ is the step function.
The key difference from the two-dimensional (2D) case considered in Sec.~\ref{sec:model-continuum} is in the overall scaling of the integrated Green's function with $\omega$. Indeed, as one can see from the results for the on-site LDOS in Fig.~\ref{fig:3D-Dirac-DOS-omega}(a), the impurity resonance peaks have a similar form albeit appear on top of the parabolic density of states (DOS) defined as
\begin{eqnarray}
\label{2D-Dirac-LDOS-GR-3D}
\nu_{0} = \frac{\omega^2}{2\pi^2 v_F^3 \Lambda^3} \theta\left(v_F \Lambda -|\omega|\right).
\end{eqnarray}
Further, as follows from Figs.~\ref{fig:3D-Dirac-DOS-omega}(b) and \ref{fig:3D-Dirac-DOS-omega}(c), the DOS becomes nonzero for $|\omega|>v_F\Lambda$ when $\mbox{Im}\left[U\right]\neq0$. Therefore, the key features of the impurity resonances and non-Hermitian defect states resemble those for 2D Dirac systems considered in Sec.~\ref{sec:model-continuum}.

\begin{figure*}[!ht]
\begin{center}
\includegraphics[width=0.32\textwidth]{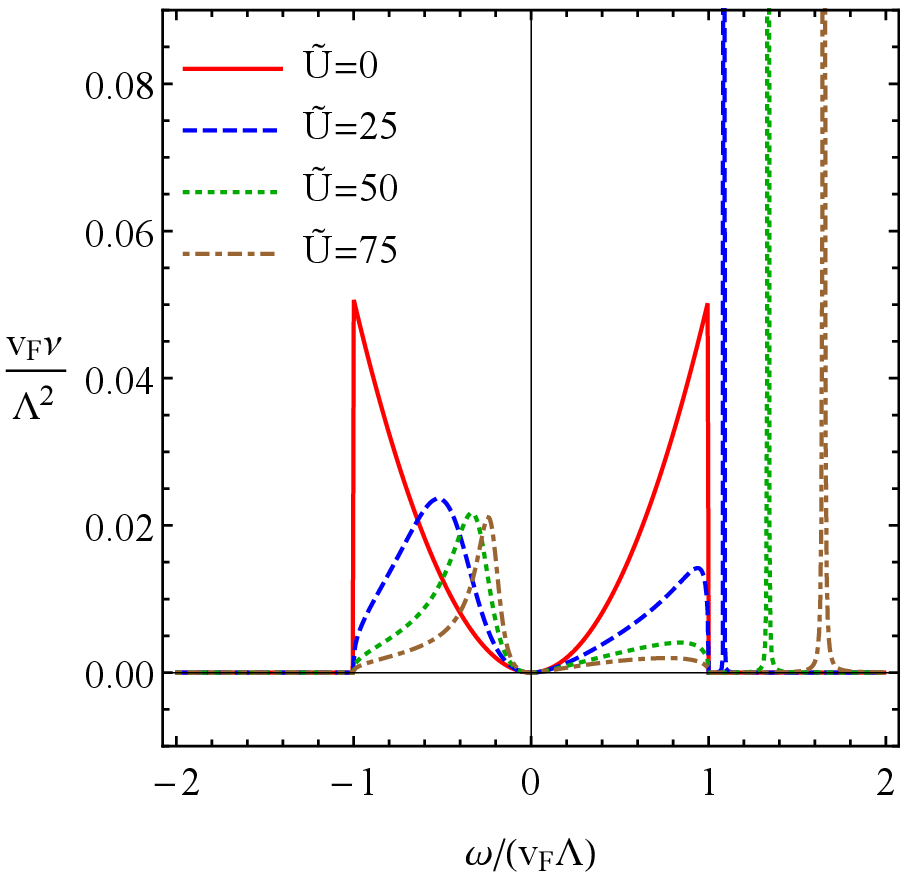}
\hfill
\includegraphics[width=0.32\textwidth]{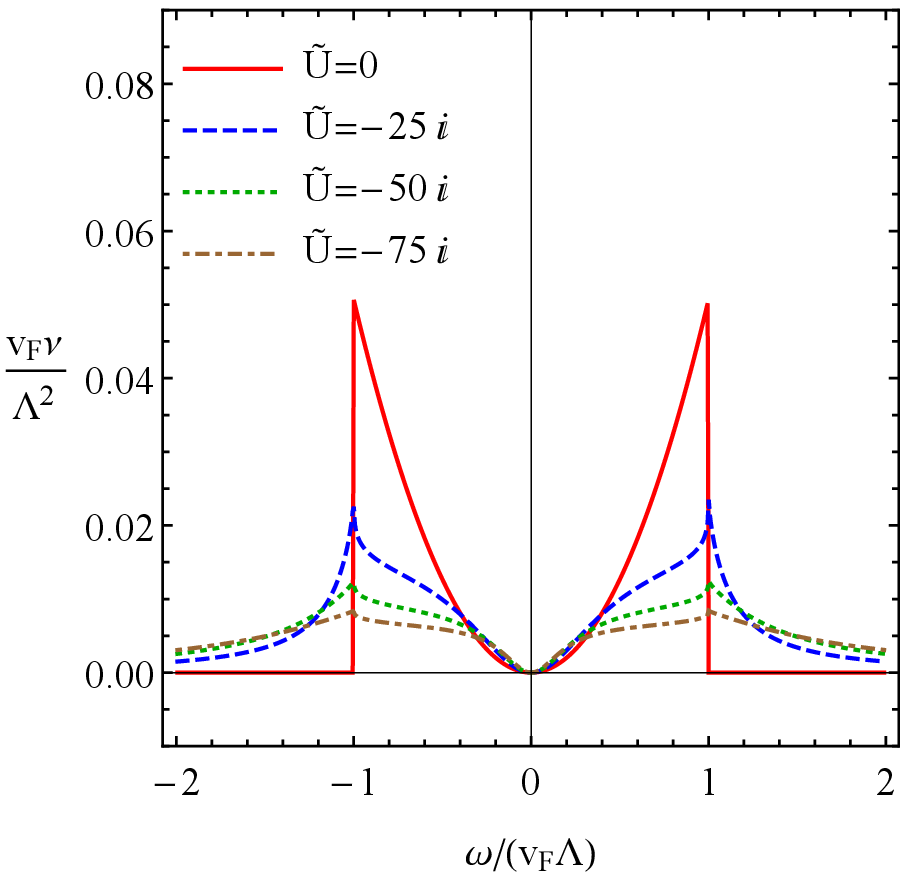}
\hfill
\includegraphics[width=0.32\textwidth]{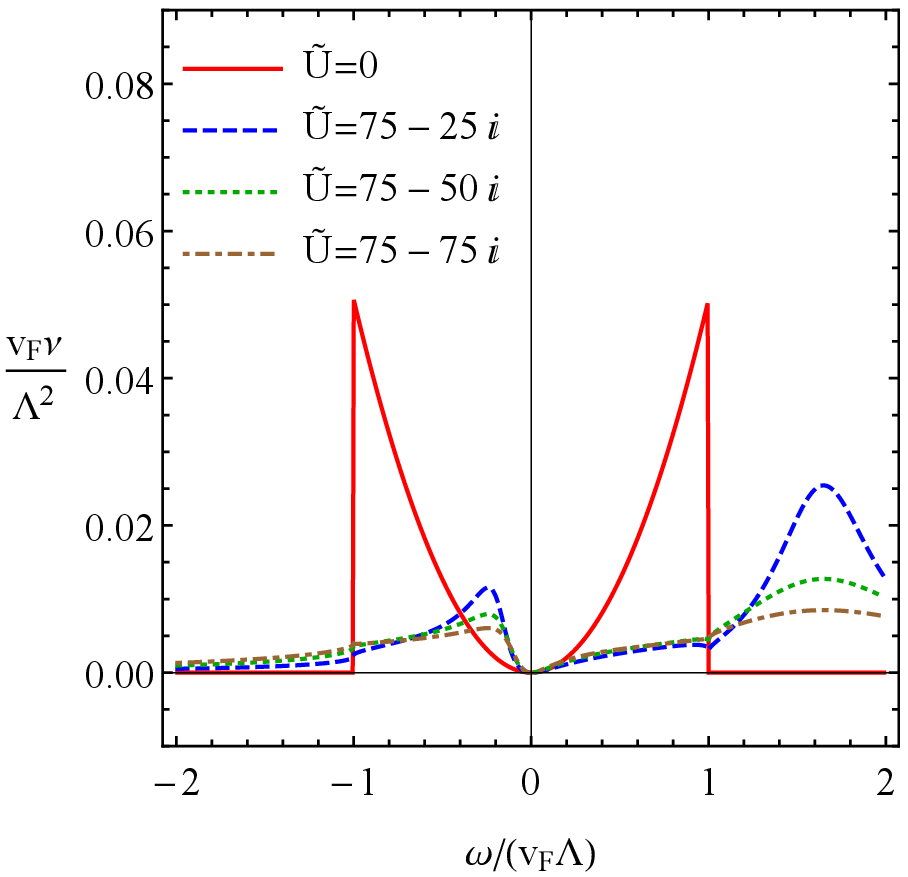}
\end{center}
\hspace{0.01\textwidth}{\small (a)}\hspace{0.32\textwidth}{\small (b)}\hspace{0.32\textwidth}{\small (c)}\\[0pt]
\caption{The DOS at the impurity site for real (panel (a)), imaginary (panel (b)), and complex (panel (c)) impurity potentials as a function of the frequency $\omega$. Red solid, blue dashed, green dotted, and brown dot-dashed lines correspond to different disorder strengths shown in the corresponding legends. In addition, $\tilde{U}=U/(v_F\Lambda)$.
}
\label{fig:3D-Dirac-DOS-omega}
\end{figure*}

As for the spatial profiles of the local DOS (LDOS), which are shown in Fig.~\ref{fig:3D-Dirac-DOS-r}, the results are also similar to those in the 2D case. The form of the peaks is, however, slightly different. In particular, while the LDOS for real potentials has a less-pronounced peak nearby the impurity (cf. Figs.~\ref{fig:2D-Dirac-DOS-r}(a) and \ref{fig:3D-Dirac-DOS-r}(a)), the peak is more noticeable for a dissipative $U$. In addition, while the magnitude of the peak is reduced by an imaginary part of the complex impurity potential in 2D, the situation is opposite in the 3D case.
It is worth noting also that while the LDOS in 2D materials is directly accessible to scanning tunneling microscopy probes, it might be hard to investigate bulk impurity resonances in 3D.

\begin{figure*}[!ht]
\begin{center}
\includegraphics[width=0.32\textwidth]{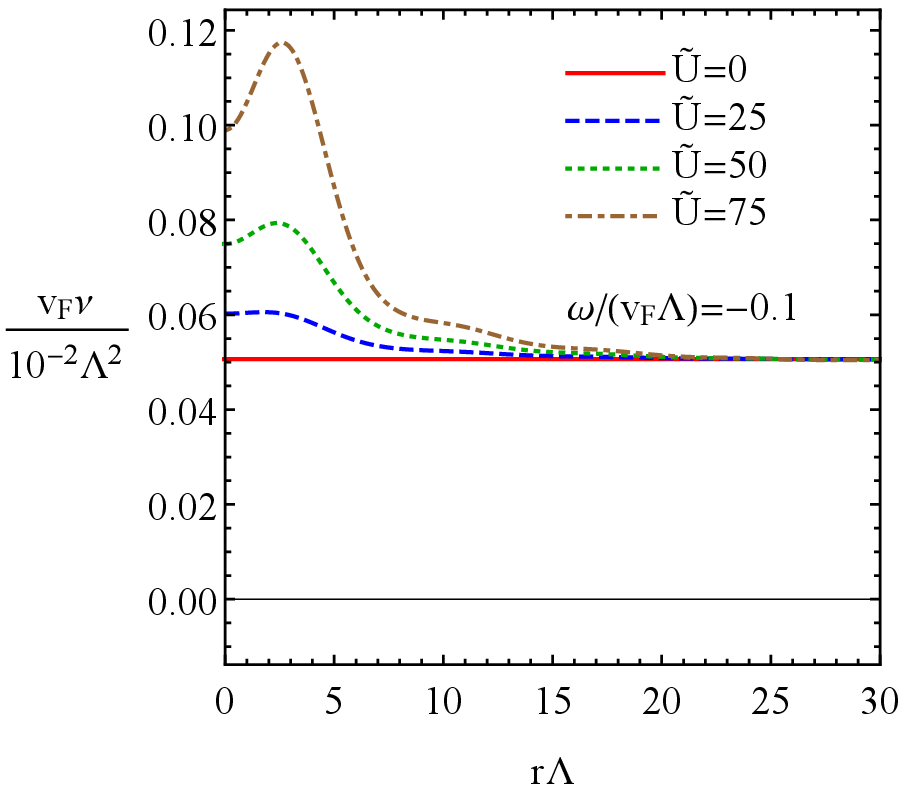}
\hfill
\includegraphics[width=0.32\textwidth]{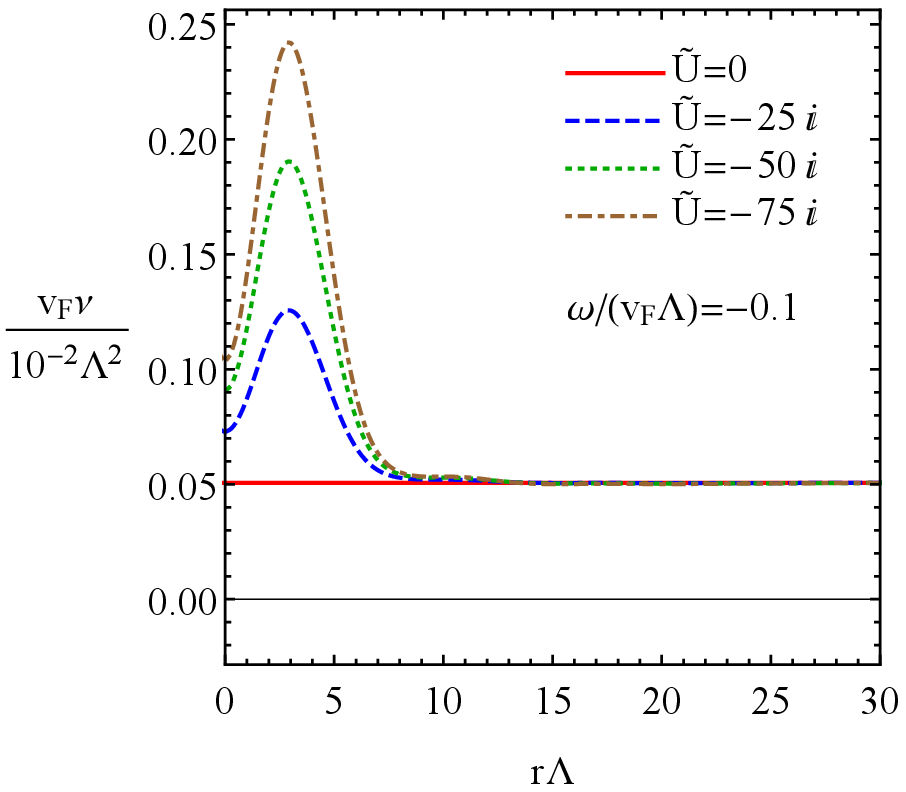}
\hfill
\includegraphics[width=0.32\textwidth]{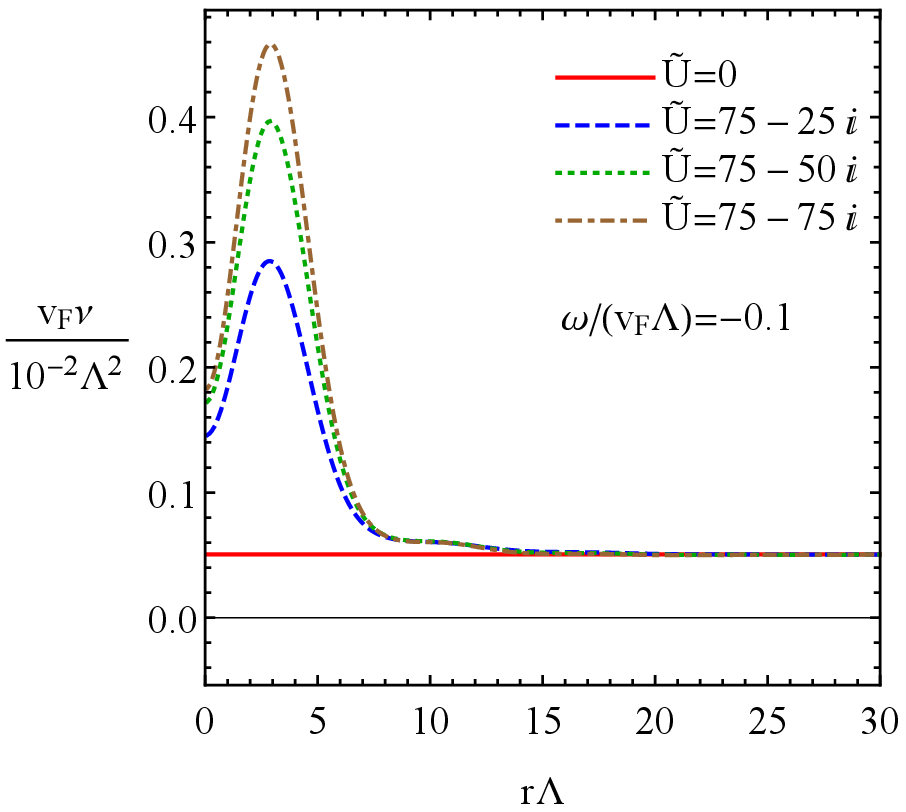}
\end{center}
\hspace{0.01\textwidth}{\small (a)}\hspace{0.32\textwidth}{\small (b)}\hspace{0.32\textwidth}{\small (c)}\\[0pt]
\caption{The LDOS for real (panel (a)), imaginary (panel (b)), and complex (panel (c)) impurity potentials as a function of the radial coordinate $r$. Red solid, blue dashed, green dotted, and brown dot-dashed lines correspond to different disorder strengths shown in the corresponding legends. In addition, we used $\omega=-0.1\,v_F\Lambda$ and $\tilde{U}=U/(v_F\Lambda)$.
}
\label{fig:3D-Dirac-DOS-r}
\end{figure*}

\section{LDOS for purely imaginary defects in the hexagonal lattice model}
\label{sec:lattice-Im}

In this section, we demonstrate the evolution of the LDOS in the hexagonal lattice model described in Sec.~\ref{sec:lattice-model}. In particular, we concentrate on the case of a purely imaginary disorder with $\mbox{Re}\left[U_A\right]=\mbox{Re}\left[U_B\right]=\mbox{Re}\left[U\right]=0$ and a negative imaginary part fo the potential, where the dependence of the LDOS on the potential strength $U$ is nonmonotonic. The results for a single impurity $\mbox{Im}\left[U_A\right]=\mbox{Im}\left[U\right]$ and $\mbox{Im}\left[U_B\right]=0$ placed at the site marked by the black dot in Fig.~\ref{fig:tb-lattice} are shown in a few panels of Fig.~\ref{fig:App-lattice-LDOS-Im}. As one can see, the LDOS first increases at the impurity site (see Figs.~\ref{fig:App-lattice-LDOS-Im}(a) and \ref{fig:App-lattice-LDOS-Im}(b)). Then, with the rise of the absolute value of the potential, the trigonal pattern in the LDOS starts to manifest (see Figs.~\ref{fig:App-lattice-LDOS-Im}(c) and \ref{fig:App-lattice-LDOS-Im}(d)). The LDOS in the latter case is similar to that for real defect potential shown in the right panel of Fig.~\ref{fig:lattice-DOS-Re}.

\begin{figure*}[ht!]
\begin{minipage}[ht]{0.45\linewidth}
\center{\includegraphics[width=1.0\linewidth]{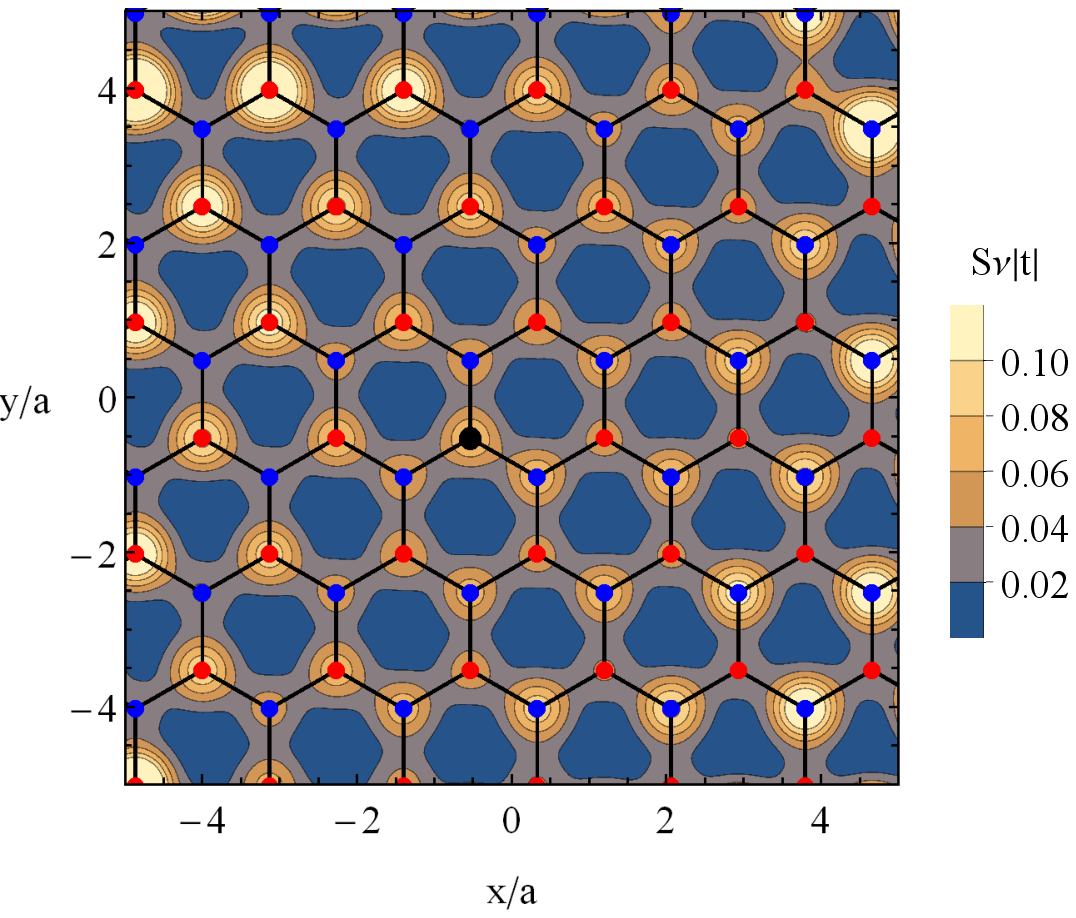} \\
{\small (a) $U=-i\,|t|$}}
\end{minipage}
\hspace{5mm}
\begin{minipage}[ht]{0.45\linewidth}
\center{\includegraphics[width=1.0\linewidth]{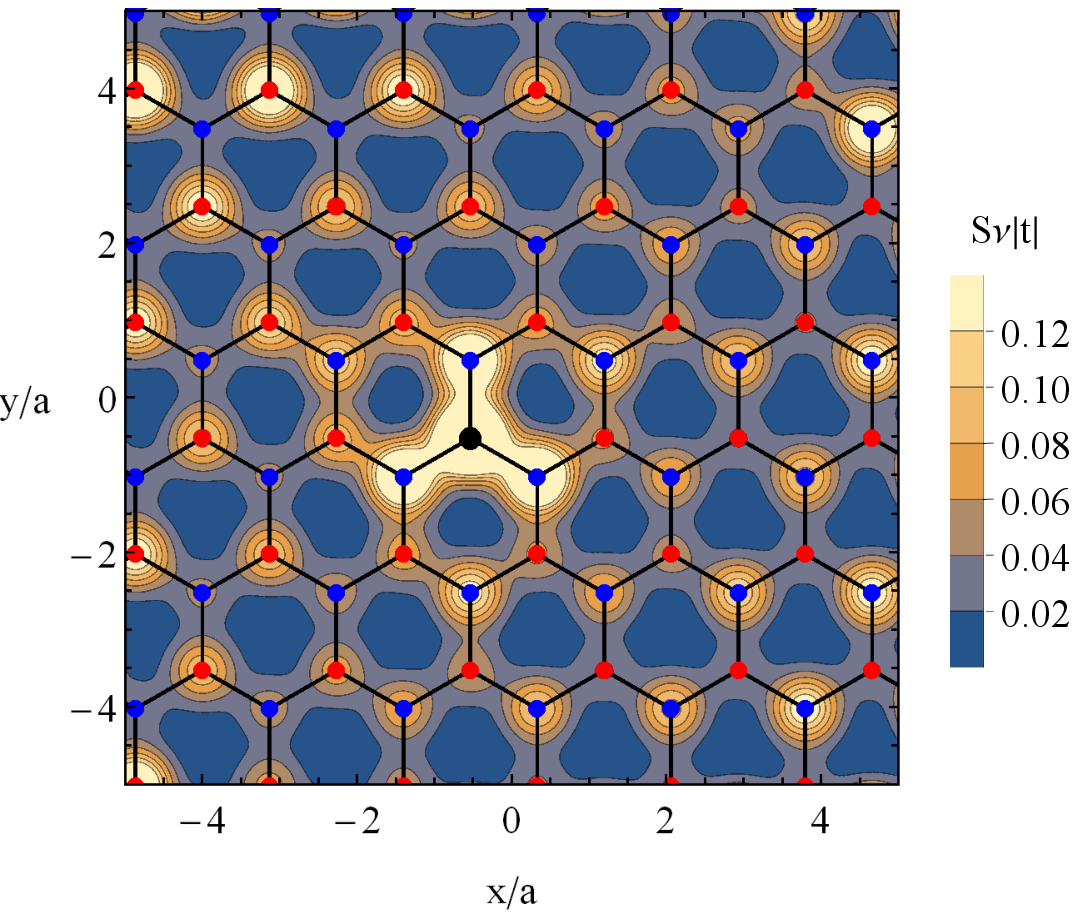} \\
{\small (b) $U=-3i\,|t|$}}
\end{minipage}
\\[5mm]
\begin{minipage}[ht]{0.45\linewidth}
\center{\includegraphics[width=1.0\linewidth]{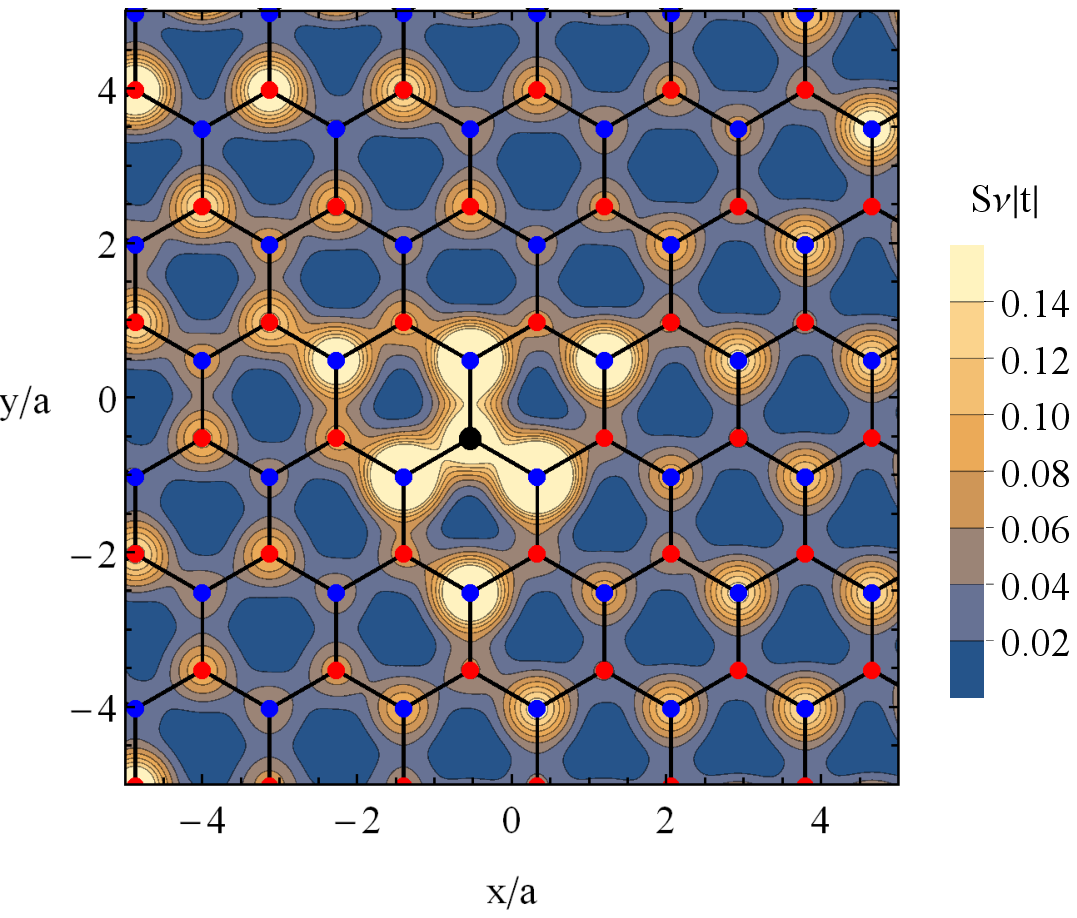} \\
{\small (c) $U=-5i\,|t|$}}
\end{minipage}
\hspace{5mm}
\begin{minipage}[ht]{0.45\linewidth}
\center{\includegraphics[width=1.0\linewidth]{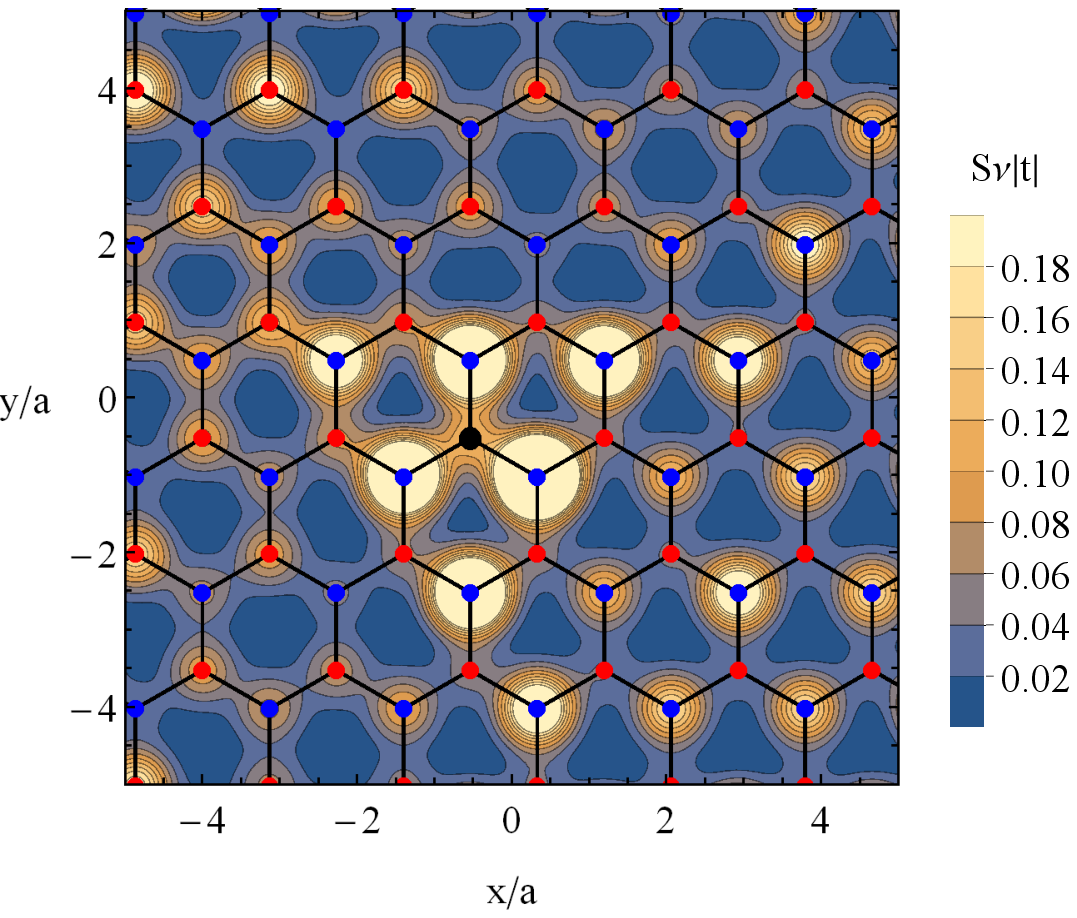} \\
{\small (d) $U=-10i\,|t|$}}
\end{minipage}
\caption{The spatial distribution of the LDOS for $\omega=0$. The position of a single impurity is marked by the black dot.
}
\label{fig:App-lattice-LDOS-Im}
\end{figure*}

The results for two defects with the same imaginary potential $\mbox{Im}\left[U_A\right]=\mbox{Im}\left[U_B\right]=\mbox{Im}\left[U\right]$ (see the black and green dots in Fig.~\ref{fig:tb-lattice}) are shown in Fig.~\ref{fig:App-lattice-LDOS-2-Im}. Similarly to the case of a single defect considered before, the dependence of the LDOS on the impurity strength is nonmonotonic. While it reaches maximum on the defect sites for small values of impurity potential, the double-trigonal pattern appears for large $|U|$.

\begin{figure*}[ht!]
\begin{minipage}[ht]{0.45\linewidth}
\center{\includegraphics[width=1.0\linewidth]{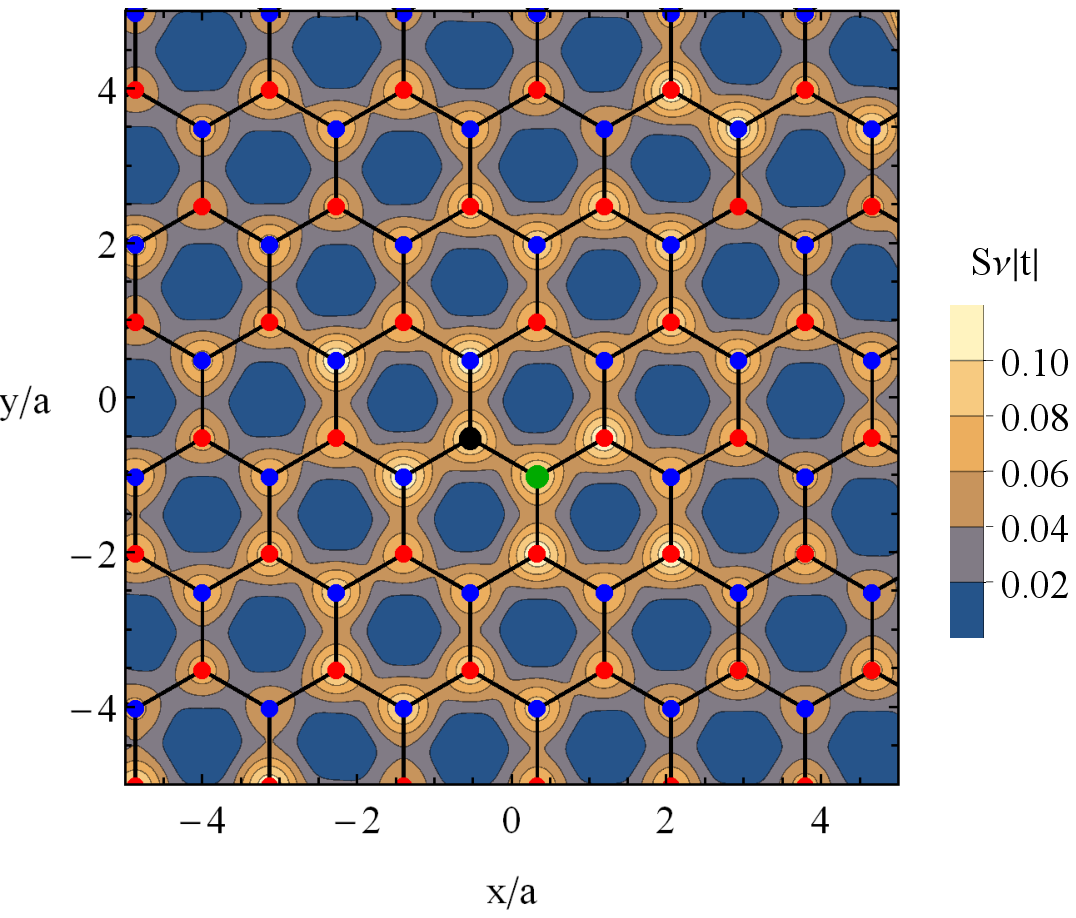} \\
{\small (a) $U=-i\,|t|$}}
\end{minipage}
\hspace{5mm}
\begin{minipage}[ht]{0.45\linewidth}
\center{\includegraphics[width=1.0\linewidth]{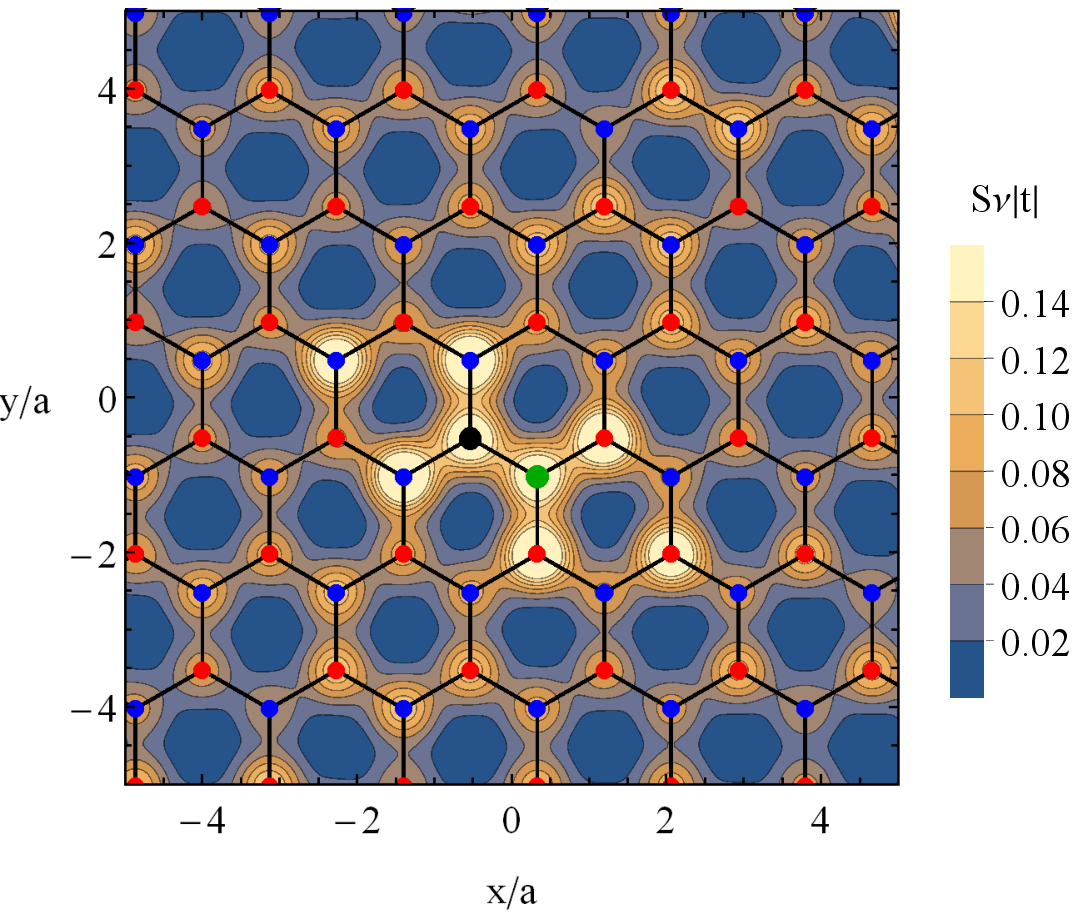} \\
{\small (b) $U=-3i\,|t|$}}
\end{minipage}
\\[5mm]
\begin{minipage}[ht]{0.45\linewidth}
\center{\includegraphics[width=1.0\linewidth]{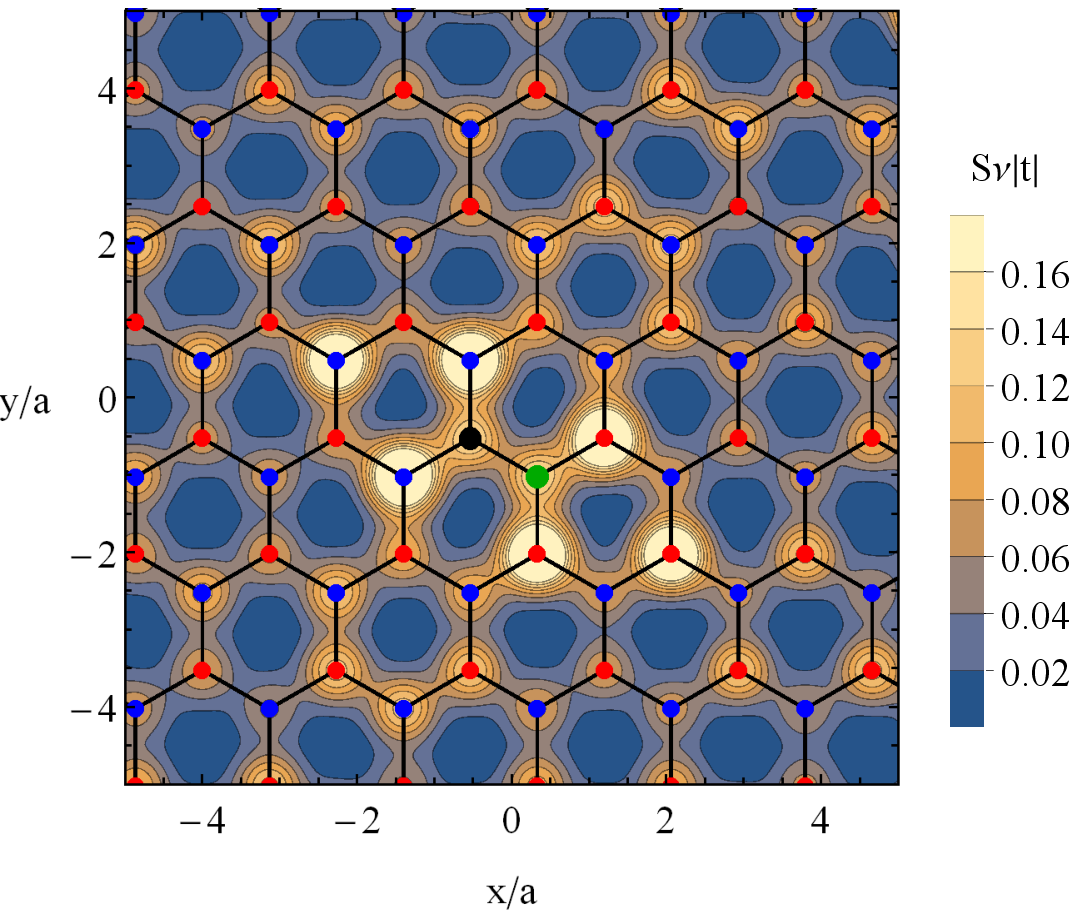} \\
{\small (c) $U=-5i\,|t|$}}
\end{minipage}
\hspace{5mm}
\begin{minipage}[ht]{0.45\linewidth}
\center{\includegraphics[width=1.0\linewidth]{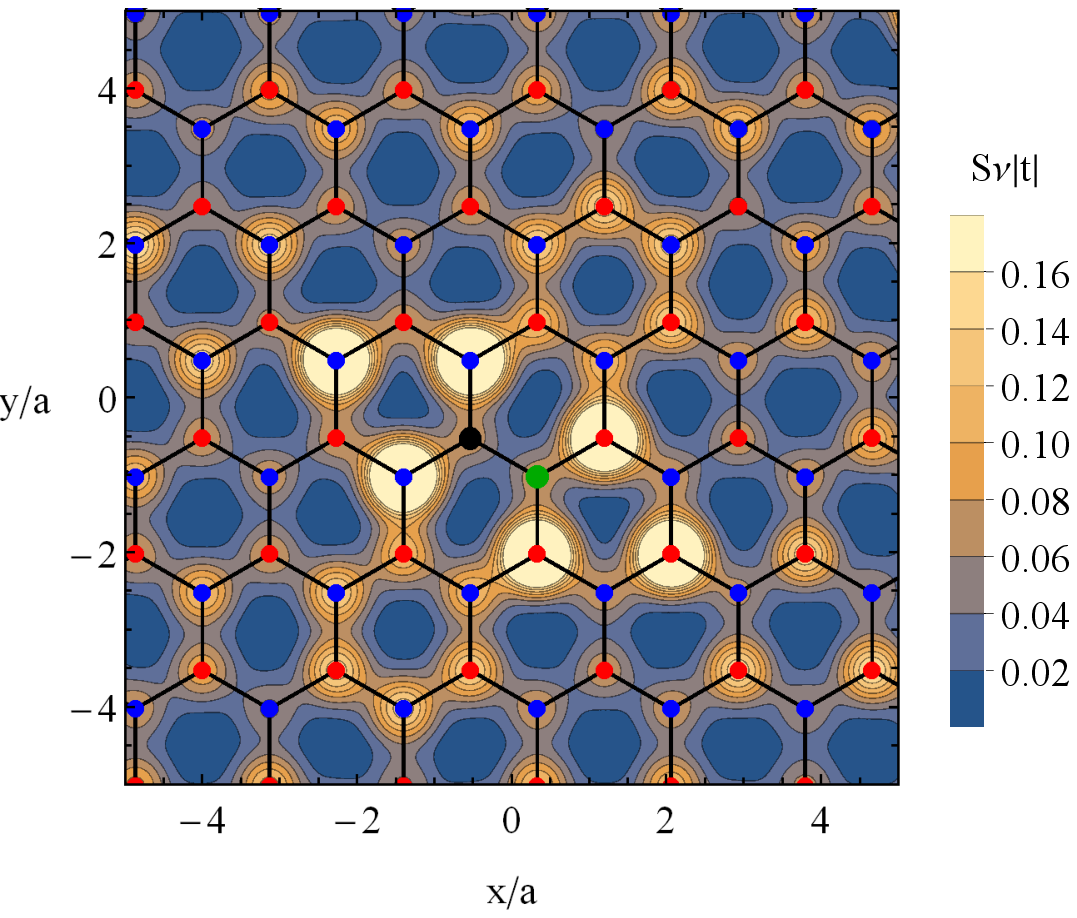} \\
{\small (d) $U=-10i\,|t|$}}
\end{minipage}
\caption{The spatial distribution of the LDOS for $\omega=0$. The position of two impurities is marked by the black and green dots.
}
\label{fig:App-lattice-LDOS-2-Im}
\end{figure*}

\end{widetext}

\end{document}